\documentclass[onecolumn,superscriptaddress,amssymb,amsmath,nobibnotes,aps,prd,showpacs,nofootinbib]{revtex4}%
\pdfoutput=1

\usepackage{graphicx}
\usepackage{epsf}
\usepackage{bm}
\usepackage{amsmath,hyperref}
\usepackage{amsfonts}
\usepackage{amssymb}
\usepackage{epstopdf}
\usepackage{natbib}%
\usepackage{rotating}
\usepackage{pdflscape}
\usepackage{adjustbox}
\usepackage[toc,page]{appendix}
\providecommand{\U}[1]{\protect\rule{.1in}{.1in}}
\newcommand{\be}{\begin{equation}}
\newcommand{\ee}{\end{equation}}

\newcommand{\mincir}{\raise
-3.truept\hbox{\rlap{\hbox{$\sim$}}\raise4.truept\hbox{$<$}\ }}
\newcommand{\magcir}{\raise
-3.truept\hbox{\rlap{\hbox{$\sim$}}\raise4.truept\hbox{$>$}\ }}

\usepackage{color}

\begin{document}

\title{Constraints On Dark Energy Models From Galaxy Clusters and Gravitational Lensing Data}

\author{Alexander Bonilla}
\email{abonilla@fisica.ufjf.br}
\affiliation{Departamento de F\'isica, Universidade Federal de Juiz de Fora, 36036-330,
Juiz de Fora, MG, Brazil}

\author{Jairo E. Castillo}
\email{jecastillo@distrital.edu.co}
\affiliation{ Facultad Tecnol\'ogica,  Unversidad Distrital Francisco Jos\'e de Caldas, Carrera 7 No. 40B - 53, Bogot\'a, Colombia}

\pacs{Published: 22 January 2018, Universe, MDPI.}

%%%%%%%%%%%%%%
\begin{abstract}

The Sunyaev--Zel'dovich (SZ) effect is a global distortion of the Cosmic Microwave Background (CMB) spectrum as a result of its interaction with a hot electron plasma in the intracluster medium of large structures gravitationally viralized such as galaxy clusters (GC). Furthermore, this~hot gas of electrons emits X-rays due to its fall in the gravitational potential well of the GC. The~analysis of SZ and X-ray data provides a method for calculating distances to GC at high redshifts. On the other hand, many galaxies and GC produce a Strong Gravitational Lens (SGL) effect, which has become a useful astrophysical tool for cosmology. We use these cosmological tests in addition to more traditional ones to constrain some alternative dark energy (DE) models, including the study of the history of cosmological expansion through the cosmographic parameters. Using Akaike and Bayesian Information Criterion, we find that the $wCDM$ and $\Lambda CDM$ models are the most favoured by the observational data. In addition, we found at low redshift a peculiar behavior of slowdown of the universe, which occurs in dynamical DE models when we use data from GC.

\end{abstract}

\maketitle

\section{Introduction}

Several authors have used the Sunyaev--Zel'dovich (SZ) effect, X-rays and Strong Gravitational Lens (SGL) data from galaxies and galaxy clusters (GC) to provide independent estimations of cosmological parameters. The combination of X-rays and the SZ data leads to two useful cosmological tests, namely angular diameter distance $d_{A}$ \cite{DeF05} and gas mass fraction $f_{gas}$ of the GC \cite{Bonamente:2006}. Both tests have been used in the literature to investigate dark energy (DE) \cite{biesiada_b} and modified gravity~(see \cite{2016arXiv161201535C} and reference therein). Additionally, the SGL observations also can be used to probe the dark matter (DM) and DE properties \cite{biesiada_b}. Therefore, to use the GC measures constitutes an independent and complementary test to probe cosmological models. Now, from the phenomenological point of view, the $\Lambda CDM$ (Cosmological constant + Cold Dark Matter) model is the most accepted to date, which predicts that the universe consists of approximately $4\%$ of baryonic matter, $26\%$ of CDM and about $70\%$ is a exotic component known as DE, which is mainly responsible for the accelerated expansion of the Universe nowadays. In the concordance model ($\Lambda CDM$), it is assumed that CDM is made up of collisionless non baryonic particles and DE is driven by cosmological constant $\Lambda$, which has an equation of state (EoS) $w=-1$. From~this perspective, the concordance model is in excellent agreement with the observations of Supernova Ia (SNIa), cosmic microwave background (CMB) anisotropies and baryonic acoustic oscillations (BAO). However, $\Lambda CDM$~model has some unresolved fundamental issues about the nature of the of DM and DE \cite{2008ARA&A..46..385F,2006astro.ph..9591A}. With respect to DE, there~are different theoretical arguments against $\Lambda$. The first is the coincidence problem, which~establishes the question of: why do the values of DE and DM density have the same order of magnitude today? Another important issue is related to ``fine tuning'' of the value of the cosmological constant to the present, which is in complete disagreement with quantum field theory and particle physics \cite{weinberg89,cop06}. In this way, several DE models with a dynamical EoS have been proposed to try to solve the so-called ``cosmological constant problem'' \cite{weinberg89,2008ARA&A..46..385F}. 

Our main aim in this paper is to impose the constraints on some well established cosmological models in the literature with the use of GC and SGL in the frame of Friedmann--Lema\^{i}tre--Robertson-- Walker (\textit{FLRW}) cosmology.

This paper is organized as follows. In Section \ref{sec:02}, we introduce the cosmological tests and the statistical analysis. In Section \ref{sec:03}, we describe cosmological models of DE, including the main results. The~history of expansion is analyzed in Section \ref{sec:04}. In Section \ref{sec:05}, we provide the summary and the~discussion. 

\section{Galaxy Clusters}
\label{sec:02}

The GC are the biggest gravitational structures in the Universe. 
They are in the transition between the linear and nonlinear regimes of the structure formation.
Gravitational lensing of background sources produced by these systems are used to infer the shape of matter distributions in the Universe. 
Nevertheless, some lensing results such as high Navarro--Frenk--White concentration parameters and the predictions of the Einstein radii 
distributions are in tension with the standard $\Lambda$CDM model \cite{bartelman}. 
Therefore, the study of GC is very important for cosmology because it offers 
information that can be used to develop cosmological tests that help to distinguish between different 
models of DE present in the literature. In what follows, we describe briefly three different data sets that will be used in the development of these cosmological tests: the GC (SZ/X-ray, $f_{gas}$) and SGL.

\subsection{Angular diameter distance using SZ/X-Ray method}
\label{sec:02.1}

The thermal SZ effect is a small distortion in cosmic microwave background (CMB) spectrum due
to the inverse Compton scattering of the CMB photons when they pass through the hot gas of electrons in GC \cite{sunyaev70,SZe}. 
This small fluctuation in CMB temperature is characterized by $\Delta T_{sz}/T_{cmb}=f(\nu,T_e)y(n_e,T_e)$, where

\begin{equation}
y(n_e,T_e) = \int_{los} n_{e} \frac{k_BT_e}{m_ec^2}\sigma_T dl,
\label{eq1:1.1} 
\end{equation}

\noindent which is known as the Compton parameter, such that $T_{cmb}=2.726$ K, $n_e$ and $T_e$ are the temperature of CMB, electron number density and temperature of the hot gas, respectively. $\sigma_T$ is the Thomson cross section, $k_B$ is the Boltzmann constant,  $m_ec^2$ is the rest mass of the electron and the integration is along the line of sight (\textit{los}). The dependence with the frequency of the thermal SZ effect is given through the term $f(\nu,T_e)$, which also introduces relativistic corrections (see \cite{Itoh et al.(1998)} for more details and \cite{Nozawa et al.(2006)} for a more recent update).

On the other hand, gas in GC can reach temperatures of $10^7-10^8$ K and densities of the order of $10^{-1}-10^{-5}$ cm$^{-3}$, so they emit high amounts of energy in X-rays. The primary emission mechanisms of X-rays for a diffuse intra-cluster medium are collisional processes such as: free--free (Bremsstrahlung), free--bound (recombination) or bound--bound (mainly emission lines), with luminosities of the order of $10^{44}$ erg/s or even higher and spatial extensions of several arcmin or larger, even at high redshift. X-rays' observations currently offer a powerful technique for building catalogs of galaxy clusters, which are very important for modern cosmology \cite{Vikhlinin et al.(2009)}.  The X-ray GC emission is given by

\begin{equation}
 S_{x} = \frac{1}{4\pi(1+z)^{4}}\int n_{e}^2 \Lambda_{eH}(\mu_e/\mu_H) dl, 
 \label{eq1:1.2}
\end{equation}

\noindent where $\Lambda_{eH}$ is the X-ray cooling function, $\mu$ is the molecular weight given by  $\mu_i=\rho/(n_im_p)$ and $z$ is the cluster redshift \cite{Bonamente:2006, DeF05}. Then, combining Equations \eqref{eq1:1.1} and \eqref{eq1:1.2} through $n_e$, we can obtain experimental cosmological distance with triaxial symmetry, given by
\vspace{12pt}
\begin{equation}
D_{c}|^{ell}_{exp} =   \frac{\Delta T_{SZ0}^{2}}{ S_{x0}} \left( \frac{m_ec^2}{k_BT_{e} } \right)^2  \frac{g(\beta)}{g(\beta/2)^2\theta_{c,proj}} \frac{  \Lambda_{eH}(\mu_e/\mu_H)}{4\pi^{3/2}f(\nu,T_e)^2T^{2}_{cmb} \sigma^2_T  (1+z)^{4}},
 \label{eq1:1.3}
\end{equation}

\noindent where $\Delta T_{SZ0}$ and  $S_{x0}$ are the central temperature decrement and the central surface brightness, respectively, which include all the physical constants and the terms resulting from the \textit{los} integration, such that $\Delta T_{SZ0}\propto d_A(z)$, $S_{x0}\propto d_A(x)$ and $d_A(z) = D_{c}|^{ell}_{exp}h^{3/4}(e_{proj}/e_1e_2)^{1/2}$, \textit{h} is a function of GC shape and orientation, $e_{proj}$ is axial ratio of the major to minor axes of the observed projected isophotes and $\theta_{c,proj}$ is the projection on the plane of the sky (\textit{pos}) (see Appendix \ref{AppxA} for some useful relationships and Table \ref{tab:Dc} for some data used in these methods). The expression in Equation \eqref{eq1:1.3} is an observational quantity that depends basically on the physical and geometrical properties of the cluster (see \cite{DeF05} for more information about the astrophysical details). That method for measuring distances is completely independent of other techniques and is valid at any redshift. We use $25$ measurements of angular diameter distances from GC obtained through SZ/X-ray method by De Filippis et al. (see Figure \ref{figure:dA}). In our analysis, we follow the standard procedure and minimize the $\chi^2$ function

\begin{equation}
\chi_{dis}^{2} (z_{i},\Theta)= \sum_{i=1}^{25} \frac{\left( D_{c}|^{ell}_{exp}(z_{i}) - d_{A}(z)\right)^{2}}{\sigma_{D_c}^{2}},
 \label{eq1:1.4}
\end{equation}

\noindent where $d_A(z)$ is the angular diameter distance in a \textit{FLRW} universe and $\sigma_{D_c}^{2}$ are the errors associated with $D_{c}|^{ell}_{exp}(z_{i})$ (see Table \ref{tab:Dc} in Appendix).  

\subsection{The gas mass fraction $f_{gas}$}
\label{sec:02.2}

Another independent cosmological technique is to derive $d_{A}$ using the gas mass fraction data from GC. 
In order to use $f_{gas}$ as a cosmological test, we need to assume that there is a proportion between the baryonic fraction of the GC and the global fraction of baryonic matter and DM. Moreover, it is necessary to assume that the baryonic fraction from clusters does not depend on the redshift \cite{Sasaki}. This~assumption is valid if one considers that these clusters are formed approximately by the same 
time.\footnote{Even though GC forms at the same time, they can have different evolution and thus different gas fractions. 
To preserve the constancy of the baryon fraction with redshift to mimic the relative cosmic abundance, GCs have to be selected among
the most massive and relaxed ones at each epoch.} (see \cite{Allen:2004cd} for more details). Thus, 
the gas mass fraction can be defined as $f_{gas}\equiv M_{gas} /M_{tot}$, where $M_{gas}$ is the X-ray's gas mass and $M_{tot}$ 
is the total gravitational mass of GC respectively. To relate $f_{gas}$ with the parameters of a particular cosmological model,  
we can write $M_{gas}$ and $M_{tot}$ in terms of $d_{A}(z)$ as follows \cite{Nesseris:2006er},

\begin{equation}
f_{gas}^{\Lambda CDM} (z) \equiv
\frac{b}{1+\alpha}\frac{\Omega_b}{\Omega_{0m}}\left(\frac{d_{A}^{\Lambda
CDM} (z)}{d_A(z)}\right)^{3/2},
 \label{eq1:1.5}
\end{equation}

\noindent where ${d_{A}}(z)$ is the angular diameter distance for a given cosmological model and $d_{A}^{\Lambda CDM}(z)$ is the angular diameter distance 
for a reference model; in this case, let us assume the $\Lambda CDM$ model. Here, $\Omega_{b}$ and $\Omega_{0m}$ are the baryonic density parameter 
and the DM density parameter, respectively. The~parameter $b$ is the depletion factor that relates the baryonic 
fraction in clusters to the mean cosmic value. The constant $\alpha$ is the ratio between optically luminous baryonic mass in 
galaxies (stellar mass) to the baryonic X-ray gas mass in intracluster medium, and its value is given by $\alpha\approx 0.19 \sqrt{h}$~\cite{Allen:2004cd}.
The~factor $h$ is the normalized Hubble constant, that is, $h=H_{0}/100\, \mathrm{km\, s}^{-1}\,\mathrm{Mpc}^{-1}$. Let us use the $f_{gas}$ measurements from $42$ GC obtained in \cite{Allen:2007ue}. The~$\chi^{2}$ is defined as
\vspace{12pt}
\begin{equation}
\begin{aligned}
&&\chi_{f_{gas}}^{2}(z_i,\Theta) = \sum_{i=1}^{42}\frac{[f_{gas}^{\Lambda
CDM}(z_i,\Theta)-f_{gas}(z_i,\Theta)]^2}{\sigma_{f_{gas}}^2} \\
&& +\left(\frac{\Omega_{b} h^{2} - 0.0214}{0.0020}\right)^2 + \left(\frac{h - 0.72}{0.08}\right)^2 + \left(\frac{b - 0.824}{0.089}\right)^2,
 \label{eq1:1.6}
 \end{aligned}
\end{equation}

\noindent where $f_{gas}$ is observational gas mass fraction data \cite{Allen:2007ue} and $\sigma_{f_{gas}}$ are the systematic errors. In the analysis, we have considered $b=0.824$ \cite{Allen:2004cd}.
 
\subsection{Gravitational lensing}
\label{sec:02.3}

The gravitational lens effect is one of the queen's tests of General Relativity. Strong gravitational lensing occurs when the light rays of a source are strongly deflected by the lens producing multiples images. 
The position of these images depend on the properties of the lens mass distribution \cite{limousin}. Because the Einstein radii, $\theta_{E}$, 
also depends on a cosmological model, the SL observations can be used as an additional method to probe the nature of the DE \cite{biesiada_b,cao}. 
In this work, we use the method that consists of comparing the ratio $\mathcal{D}$ of angular diameter distances between lens and source,  
$d_{A}(z_{l},z_{s})$, and between observer and lens, $d_{A}(0,z_{s})$, with its observable counterpart $\mathcal{D}^{obs}$ given by

\begin{equation}
\mathcal{D} (z_{l},z_{s}) = \frac{d_{A}(z_{l},z_{s})}{d_{A}(0,z_{s})} = 
\frac{\int_{z_{l}}^{z_{s}} dz'/E(z',\Theta)}{\int_{0}^{z_{s}} dz'/E(z',\Theta)},
 \label{eq1:1.7}
\end{equation}

\begin{equation}
\mathcal{D}^{obs} = \frac{c^{2} \theta_{E}}{4\pi \sigma_{SIS}^{2}}, 
\label{eq1:1.8}
\end{equation}

\noindent where $\sigma_{SIS}$ is the Singular Isothermal Sphere (SIS) velocity dispersion and $E (z,\Theta) \equiv H(z,\Theta)/H_{0}$, $H(z,\Theta)$ being the Hubble function. In order to put constraints on cosmological parameters through $E (z,\Theta)$, the Einstein radius $\theta_{E}$ and the dispersion velocity $\sigma_{SIS}$ (exactly its central velocity dispersion~$\sigma_0$) must be obtained by astrometric and spectroscopic means, respectively. In the first case, it depends on the lens modelling (either SIS, Singular Isothermal Ellipsoid (SIE) or Navarro--Frenk--White density profiles). In the second case, the velocity dispersion $\sigma_{SIS}$ of the mass distribution and the observed stellar velocity dispersion $\sigma_0$ need not be the same, since the halos of DM can have a greater speed of dispersion than the visible stars  \cite{1996AAS...189.4104W}. These effects can be taken into account through the following relationship  $\sigma_{SIS}=f_E \sigma_0$, where the parameter $f_E$ emulates the systematic errors in the RMS due to the difference between $\sigma_{SIS}$ and $\sigma_0$; the rms error caused by assuming the SIS model, since the observed image separation does not directly correspond to $\theta_{E}$ and softened SIS potentials which tend to decrease the typical image separations \cite{1996astro.ph..6001N}. In the present work we assume the best-fit reported in~\cite{cao} (and~references therein), where $f_E\approx 1$, which has been properly marginalized. On the other hand, GC can also act as sources to produce strong gravitational lensing showing giant arcs around GC. This phenomenon can be used to constrain the astrophysical properties of the cluster (projected mass) and cosmology \cite{2004MNRAS.354.1255S}. If we assume the condition of hydrostatic equilibrium \footnote{The pressure gradient force of an isothermal gas with temperature $T_X$ is balanced by the gravity in GC.} and an approximation of spherical symmetry \footnote{Specifically, a hydrostatic isothermal spherical symmetric $\beta-model$.} \cite{1976A&A....49..137C}, then a theoretical surface density can be described as

\begin{equation}
\Sigma_{th} = \frac{3}{2G\mu m_p}\frac{k_BT_X\beta_X}{d_A(0,z_l)\theta_c},
 \label{eq1:1.9}
\end{equation}

\noindent where $k_B$, $m_p$, $\mu=0.6$ and $\beta_X$ are, respectively, the Boltzmann constant, the proton mass, the mean molecular weight and the slope of the $\beta-model$ \cite{2002ARA&A..40..539R}. Although the hydrostatic equilibrium and isothermal hypotheses are very strong, the total mass density obtained under such assumption may lead to good estimates, even in dynamically active GC with irregular morphologies in X-rays. Then, combining this with the critical surface mass density for lensing $\Sigma_{obs}$ \cite{1992grle.book.....S}, We can get a Hubble constant independent ratio as

\begin{equation}
\mathcal{D}^{obs} =  \frac{d_{A}(z_{l},z_{s})}{d_{A}(0,z_{s})} =  \frac{\mu m_pc^2}{6\pi}\frac{1}{k_BT_X\beta_X}\sqrt{\theta_t^2+\theta_c^2},
 \label{eq1:1.10}
\end{equation}

\noindent where the parameters   $T_X$, $\beta_X$ and $\theta_c$ can be obtained from X-ray observational data. The position of tangential critical curve $\theta_t=\epsilon\theta_{arc}$, where $\theta_{arc}$ is the observational arc position and $\epsilon = (1/\sqrt{1.2})\pm 0.04$ quantifies the slight difference with arc radius angle (See \cite{1999PASJ...51...91O,2011RAA....11..776Y} for more details about the priors and 10 galaxy clusters used as sample).

In the present work we use a sample of $80$ strong lensing systems by \cite{cao}, which contains $70$ data points from SLACS and LSD and $10$ data points from GC. Again, the fit of the theoretical models to strong lensing observations can be found by the minimization of
\begin{equation}
\chi_{SL}^2 = \sum_{i=1}^{80} \frac{ \left( \mathcal{D}_{i}^{obs} - \mathcal{D}_{i}^{th} \right)^2  }{ \sigma_{\mathcal{D},i}^{2} },
 \label{eq1:1.11}
\end{equation}

\noindent where the sum is over the sample and $\sigma_{\mathcal{D},i}^{2}$ denotes the variance of $\mathcal{D}_{i}^{obs}$.

Additionally to these data sets defined in the Sections \ref{sec:02.1}--\ref{sec:02.3}, we will use 580 Supernovae data (SNIa) from Union 2.1 \cite{Suzuki et al.(2012)}, the CMB shift parameter \cite{2013PhRvD..88d3522W} (Planck 2013), as well as data from BAO (BOSS, WiggleZ, SDSS, 6dFGS) observations, adopting the three measurements of 
$A(z)$ obtained from~\cite{Anderson et al.(2012),Blake et al.(2011)}, and using the covariance among these data given in \cite{Shi:2012ma}. Each $\chi^2$ function is constructed in a way analogous to the other tests considered above (see Appendix \ref{AppxB}).

\subsection{Statistic analysis}

The procedure of finding a set of parameters for a given statistic is known as Maximum likelihood $\mathcal{L}_{max}$, that is, given a probability distribution this is maximum for the corresponding data set. The~maximum likelihood estimate for the best fit parameters $p_{i}^{m}$ is given by

\begin{equation}
 \mathcal{L}_{max}(p_{i}^{m})= exp \left[ -\frac{1}{2}{\chi_{min}^{2}(p_{i}^{m})} \right].
  \label{eq4:4.1}
\end{equation}

If $\mathcal{L}_{max}(p_{i}^{m})$ has a Gaussian errors distribution, then $\chi_{min}^{2}(p_{i}^{m})=-2 \ln \mathcal{L}_{max}(p_{i}^{m})$, which is our case \cite{2010arXiv1012.3754A}. In order to find the best values of the free parameters of the model, let us consider
\begin{equation}
\chi_{total}^{2}=\chi^{2}_{SNIa}+\chi^{2}_{CMB} +\chi^{2}_{BAO}+ \chi^{2}_{d_{A}}+\chi^{2}_{f_{gas}}+\chi^{2}_{SGL}.
\end{equation}

The Fisher matrix is used in the analysis of the constraint of cosmological models for different observational test \cite{Albrecht:2009ct,2012JCAP...09..009W}. It contains the Gaussian uncertainties $\sigma_i^2$ of the different parameters $p_{i}^{m}$. Given the best fit $\chi_{min}^{2}(p_i^{m}, \sigma_i^2)$ for a set of parameters $p_{i}^{m}$ with uncertainties $\sigma_i^2$, the Fisher matrix is 

\begin{equation}
F_{ij}=\frac{1}{2}\frac{\partial ^2 \chi_{min}^{2}}{\partial p_i^{m} \partial p_j^{m}}
\label{eq4:4.4} 
\end{equation}

\noindent for each model \textit{m}. The inverse of the Fisher matrix provides an estimate of the covariance matrix through $\left[ C_{cov} \right] = \left[ F \right]^{-1}$. Its diagonal elements are the squares of the uncertainties in each parameter marginalizing over the others, while the off-diagonal terms yield the correlation coefficients between parameters. The uncertainties obtained in the propagation of errors are given by $\sigma_i = \sqrt{Diag \left[ C_{cov} \right]_{ij} }$. Notice that the marginalized uncertainty is always greater than (or at most equal to) the non-marginalized one: marginalization can’t decrease the error, and only has no effect if all other parameters are uncorrelated with it \footnote{For an unbiased estimator, If all the parameters are assumed to be known (in other words, if we don’t marginalize over any other parameters), then the minimal expected error is  $\sigma_i = 1/\sqrt{F_{ij}}$.}. Previously known uncertainties on the parameters, known~as priors, can be trivially added to the calculated Fisher matrix. This is manifestly the case for us: a lot of standard cosmological datasets provide priors on our previously defined cosmological parameters. The analysis with the Fisher matrix is used to evaluate the errors on the best-fit parameters.

In our results, let us consider different cosmological models. Thus, a way to quantify which model best fit the data is  consider a Bayesian comparison. We adopted the Akaike and Bayesian information criterion (AIC and BIC, respectively), which allows us to compare cosmological models with different degrees of freedom, with respect to the observational evidence and the set of parameters~\cite{Akaike:74,Schwarz:78}. The~AIC and BIC can be calculated as
\begin{equation}
AIC=-2 \ln \mathcal{L}_{max} +2d,
\end{equation}
\begin{equation}
BIC=-2 \ln \mathcal{L}_{max} + d \ln N, 
\end{equation}

\noindent where $\mathcal{L}_{max}$ is the maximum likelihood of the model under consideration ($\mathcal{L}_{max}= exp \left[ -\frac{1}{2}{\chi_{min}^{2}} \right]$), $d$ is the number of parameters and $N$ the number of data points. The BIC imposes a strict penalty against extra parameters for any set with \textit{N} data. The prefered model is that which  minimizes the AIC and BIC.  However, the absolute values of them are not of interest,  only the relative values between the different models  \cite{Liddle:04}. Therefore, the ``strength of evidence'' can be characterized in the form $\Delta AIC=AIC_{i}-AIC_{min}$, 
$\Delta BIC=BIC_{i}-BIC_{min}$, where the subindex $i$ refers to value of $AIC$ ($BIC$) for model $i$ and $AIC_{min}$ ($BIC_{min}$) is the minimum value of $AIC$ ($BIC$) among all the models~\cite{Burnham:03}. We~give the judgements for both critera as follows: (i) If $\Delta AIC (\Delta BIC) \leq 2$,  then the concerned model has substantial support with respect to the reference model (i.e., it has evidence to be a good cosmological model),
(ii) if $4 \leq \Delta AIC (\Delta BIC) \leq 7$, it is an indication for less support with respect to the reference model, and, finally,  (iii) if $\Delta AIC (\Delta BIC) \geq 10$, then the model has no observational support. Thus, if we have a set of models of DE, first we should estimate the best fit $\chi^{2}$ and then we can apply the $AIC$ and $BIC$ to identify which model is the preferred one by the observations. We also apply the reduced chi-square to see how well the model fit the data, which is defined as $\chi_{red}^{2}=\chi_{min}^{2}/\nu$, where $\nu$ is the degrees of freedom usually given by $N-d$. Then, the total number of data points is: SNIa (580), CMB (3), BAO (7), $d_{A}$~(25), $f_{gas}$ (42), SGL (80), so N = 737. Priors used in the present analysis are standard and the most conservative possible and combining GC data with independent constraints from CMB, BAO and SNIa removes the need of priors for $\Omega_b$, and \textit{h} leads to tighter constraints over $\Omega_m$, $\Omega_k$ and the parameters that characterize the DE density for different cosmological models. On the other hand, SGL offers a great opportunity to constrain DE features without prior assumptions on the fiducial cosmology. In what follows, we present our main results.

 \section{Dark energy models and Results}
 \label{sec:03}

In order to put constraints on DE models using GC ($d_{A}$, $f_{gas}$) and SGL, we need to compute the angular diameter distance of the model and compare it with observational data. In addition, to~investigate whether a cosmological model can predict an accelerated expansion phase of the Universe, we must study the behavior of the deceleration parameter $q(z)$. The angular diameter distance for a FLRW universe, from a source at redshift $z$, is given by
  
\begin{equation}
d_{A} (z,\Theta) =  \frac{3000\, \mathrm h^{-1}}{(1+z)} \frac{1}{\sqrt{\mid \Omega_{k} \mid}} 
\sin \varsigma \left( \int_{0}^{z} \frac{\sqrt{\mid \Omega_{k} \mid}}{E(z,\Theta)}dz\right), 
\end{equation}
  
\noindent where h is dimensionless Hubble parameter ($H_{0}$ = h 100 $ \mathrm{km\,s}^{-1}\, \mathrm{Mpc}^{-1}$) and the function $\sin \varsigma(x)$ is defined such that it can be $\sinh(x)$ for $\Omega_{k} >0$, $\sin(x)$  for $\Omega_{k} <0$ and $x$ for $\Omega_{k} =0$ \cite{Hogg:1999ad}.  In the standard FLRW cosmology, the expansion rate as a function of the scale factor $H(a)$ is given by the Friedmann equation as
 \begin{equation}
 E^2(a,\Omega_i) = \Omega_{r}a^{-4} + \Omega_{m}a^{-3} + \Omega_{k}a^{-2} + \Omega_{X} e^{3\int_a^1 \frac{da'}{a'} \left( 1+w(a')\right) },
 \label{eq2:2.2}
 \end{equation}
  
\noindent where $H(a)/H_0=E(a,\Omega_i)$, $H_0$ is the current value of the expansion rate and the scale factor is related to redshift as $1+z=a^{-1}$, such that $a_0=1$ at present.~In Equation  (\ref{eq2:2.2}), $\Omega_i$ is a dimensionless energy densities relative to critical ($\rho_{cri}= 3 H_0^2 / 8\pi G$) in the form of the \textit{i}th component  of the fluid density of: radiation ($\Omega_{r}$), matter ($\Omega_{m}$), curvature ($\Omega_{k}$) and DE ($\Omega_{X}$). $\Omega_{r0}(h)=\Omega_{\gamma} (h) (1 + 0.2271 N_{eff})$, where~$\Omega_{\gamma} (h) = 2.469\times10^{-5}\, \mathrm h^{-2}$ % we add "\times" between "2.469" and "10^{-5}", please confirm.
is the density of photons and $N_{eff}=3.046$ is the effective number of neutrino species \cite{2014A&A...571A..16P}. $\omega (a)=p(a)/\rho (a)$ is the EoS for DE, where $p(a)$ is the fluid pressure. This~EoS divides our models into two cases: when the energy density of the fluid is constant and the energy density of the fluid is dynamic. In all cosmological models, $\Omega_k$ is a free parameter. A vector of parameters is considered for each DE model as $\Theta_{i}^{model}= \left\lbrace \theta_i, \Omega_i \right\rbrace $, where $\theta_i = \left\lbrace h, \Omega_b \right\rbrace$ for the analysis of the present work. 

\subsection{$\Lambda CDM$}
 \label{sec:03.1}
  
Our analysis starts with standard cosmological model, where DE density is provided by the cosmological constant $\Lambda$. 
The expansion rate within $\Lambda CDM$ context is given by
 
\begin{equation}
E^{2}(z,\Theta) = \Omega_{r}(1+z)^{4} + \Omega_{m}(1+z)^{3} + \Omega_{k}(1+z)^{2} + \Omega_{\Lambda},
\label{eq2:2.3}
\end{equation} 

\noindent where $\Omega_{r}$, $\Omega_{m}$ and $\Omega_{\Lambda} = 1-\Omega_{m}-\Omega_{k}-\Omega_{r}$, are the density parameters for radiation, matter and DE component, respectively. Here, the free parameter vector is $\Theta=\left\lbrace h,\Omega_b, \Omega_{m},\Omega_{k}\right\rbrace$. We find the best fit of parameters at 1$\sigma$ confidence level (CL), whose results are shown in Table \ref{tab:LCDM}.
  
\begin{table}[htb]
\centering
\begin{tabular}{lcc}
\hline 
\textit{Parameter} & \textit{CMB+BAO+SNIa} & \textit{CMB+BAO+SNIa+$d_A$+$f_{gass}$+SGL} \\
\hline 
\hline 
h                         & $0.6858\pm 0.0095$  & $0.7063\pm 0.0067$ \\ 
$\Omega_m$     & $0.2981\pm 0.0093$  & $0.2839\pm 0.0046$ \\
$\Omega_k$      & $-0.0011\pm 0.0031$ & $ 0.0048\pm 0.0024$ \\ 
$\Omega_b$      & $0.0475\pm 0.0014$  & $0.04411\pm 0.00099$ \\ 
$\chi_{min}^{2}$ & 565.686                      & 777.256 \\
\hline 
\end{tabular} 
\caption{Summary of the best fit values for $\Lambda CDM$ model.}
\label{tab:LCDM}
\end{table}
 
In Table \ref{tab:LCDM}, we can see the impact of adding the GC and SGL tests to the more traditional ones (CMB + BAO + SNIa), which evidently improves the constraints on the parameters of the model (see~Figure \ref{figure:ConDia}). 
 
\subsection{$wCDM$ model}
\label{sec:03.2}
 
The most simple extension of the $\Lambda CDM$ model is to consider that the EoS remains constant but its value can be $w\neq-1$. In this case, the expansion rate for FLRW cosmology reads as

\begin{eqnarray}
E^{2}(z,\Theta)&=&\Omega_{r}(1+z)^{4} + \Omega_{m}(1+z)^{3} + \nonumber\\ 
&&\Omega_{k}(1+z)^{2} + \Omega_{X} (1+z)^{3(1+w)},
\label{eq2:2.4}
\end{eqnarray}

\noindent where $\Omega_{X}=(1-\Omega_{m}-\Omega_{k}-\Omega_{r})$.  In this model, the set of free parameters is $\Theta=\left\lbrace h,\Omega_b,\Omega_{k}, \Omega_{m},w \right\rbrace$. As in the case of $\Lambda$CDM model, first we estimate the best fit values using the data from $SNIa+CMB+BAO$ and then, we use the full data set $SNIa+CMB +BAO+ d_{A}+f_{gas}+SGL$. The best fit values at 1$\sigma$ CL for this case is shown in the Table \ref{tab:WCDM}.
 
 \begin{table}[htb]
 \centering
 \begin{tabular}{lcc}
 \hline 
\textit{Parameter} & \textit{CMB+BAO+SNIa} & \textit{CMB+BAO+SNIa+$d_A$+$f_{gass}$+SGL} \\
 \hline 
 \hline 
h                         & $0.6897\pm 0.0098$  & $0.7080\pm 0.0070$ \\  
$\Omega_m$     & $0.2964\pm 0.0093$  & $0.2839 \pm 0.0049$ \\  
$\Omega_k$      & $-0.0028\pm 0.0033$ & $0.0007\pm 0.0028$ \\ 
$\omega$           & $-1.057\pm 0.041$     & $-1.086 \pm 0.038$ \\
$\Omega_b$      & $0.0468\pm 0.0014$  & $0.0437\pm 0.0010$ \\
$\chi_{min}^{2}$ & 563.953 & 772.283 \\
\hline 
\end{tabular} 
\caption{Summary of the best fit values for wCDM model}
\label{tab:WCDM}
\end{table}

Notice that in both cases the EoS has a phantom behavior and the standard model is excluded at least up to $2\sigma$ CL (see Figure \ref{figure:ConDia}). As the case of $\Lambda$CDM model, the curvature parameter changes from negative to positive  (Table \ref{tab:WCDM}).

\subsection{Chevalier-Polarski-Linder model}\label{sec:03.3}
 
Another simple extension to the $\Lambda CDM$ model is to allow for the EoS of the DE varies with the redshift.
Several parameterizations have been considered in the literature. Here, let us consider the popular Chevallier--Polarski--Linder (CPL) 
model \cite{Chevallier:2000qy,Linder:2003nc}
\begin{equation}
w(z) = w_{0} + w_{1} \frac{z}{1+z},
\label{eq:wcpl}
\end{equation}
where $w_0$ is the value of the DE state equation at the present and the parameter $w_{1}$ evaluates the dynamic character of DE.
The FLRW $E(z)$ for CPL parametrization is given by 
\begin{equation}
 E^{2}(z) = \Omega_{r}(1+z)^{4} + \Omega_{k} (1+z)^2+\Omega_{m}(1+z)^{3} + 
  \Omega_{X} X(z),
  \label{eq2:2.6}
\end{equation}
where $\Omega_{X} =  \left( 1 - \Omega_{k} - \Omega_{m} -\Omega_{r} \right)$ and
\begin{equation}
X(z)=(1+z)^{3(1 + w_{0} + w_{1})} \exp \left[ - \frac{3w_{1}z}{1+z}\right].
\label{eq2:2.7}
\end{equation}
 
The free parameters are $\Theta=\left\lbrace h,\Omega_b,\Omega_{k},\Omega_{m}, w_{0}, w_{1} \right\rbrace$. The best fit values at 1$\sigma$ CL using $CMB+BAO+SNIa$ and full data set are summarized in Table \ref{tab:CPL}.
 
\begin{table}[htb]
\centering
\begin{tabular}{lcc}
\hline 
\textit{Parameter} & \textit{CMB+BAO+SNIa} & \textit{CMB+BAO+SNIa+$d_A$+$f_{gass}$+SGL} \\
\hline 
\hline 
h                         & $0.688\pm 0.011$      & $0.7073\pm 0.0075$ \\
$\Omega_m$     & $0.297\pm 0.010$      & $0.2856\pm 0.0059$ \\ 
$\Omega_k$      & $-0.0054\pm 0.0055$ & $-0.0017\pm 0.0040$ \\ 
$\omega_0$       & $-0.97\pm 0.19$         & $-0.97\pm 0.15$ \\
$\omega_1$       & $-0.50\pm 1.13$         & $-0.71\pm 0.95$\\
$\Omega_b$      & $0.0470\pm 0.0015$   & $0.0439\pm 0.0011$\\ 
$\chi_{min}^{2}$ & 563.854                       & 771.481 \\
\hline 
\end{tabular} 
\caption{Summary of the best fit values for CPL model}
\label{tab:CPL}
\end{table}
 
We can see that, for both the combined analyses, the CPL model allows a quintessential DE at the current time. The curvature parameter $\Omega_k$ remains negative.  The standard model remains within the $1\sigma$ and $2\sigma$ of CL for the present analysis (see Figure \ref{figure:ConDia}).
 
\subsection{Interacting Dark Energy model}\label{sec:03.4}

Cosmological models, where DM and DE are non minimally coupled throughout the evolution history of the universe, have been considered to solve the problem of the cosmic coincidence as well as the problem of the cosmological constant (models where DM interacts with vacuum energy or Interacting Dark Energy (IDE) Models---see~\mbox{\cite{2015IJMPD..2430007B,2016RPPh...79i6901W}}  for general review). It has recently been shown that the current observational data can favor the late-time interaction  in the dark sector \cite{2014PhRvL.113r1301S,2015arXiv150602518R,2016PhRvD..94b3508N,2017arXiv170202143K,2017MNRAS.466.3497S,2016PhRvD..94l3511K}. In~general, we assume that DM and DE interact via a coupling function $Q$ given by
\begin{eqnarray}
 \dot{\rho_{m}} &+& 3H\rho_{m} = Q \rho_{m} \nonumber,\\
 \dot{\rho_{x}} &+& 3H\left(1 + w_{x}\right)\rho_{x}= -Q \rho_{m},
 \label{eq2:2.8}
\end{eqnarray}

\noindent where $\rho_{m}$ and $\rho_{x}$ are the DM and DE density, respectively, with $w_{x}$ the EoS for DE. Here,~\mbox{$Q = \delta H$} characterizes the strength of the interacting through the dimensionless coupling term $\delta$, which~establishes a transfer of energy from DE to DM for $\delta>0$, whereas for $\delta<0$ the energy transfer is the opposite. This model was originally introduced in \cite{2005CQGra..22..283W}, and then investigated in various contexts~\cite{2009PhRvD..79l7302C,2014GReGr..46.1820N,2017arXiv170509278Y}. The expansion rate of the Universe for this model is given by

\begin{eqnarray}
 E^{2}(z,\Theta) &=& \Omega_{r}(1+z)^{4} + \Omega_{k}(1\,+\,z)^{2} +  \Omega_{m}\Psi (z)\nonumber \\
  &&+  \Omega_{X} (1+z)^{3(1 + w_{x})},
\label{eq2:2.9}
\end{eqnarray}
\noindent where  $\Omega_{X} = (1 - \Omega_{m} - \Omega_{k}-\Omega_{r})$ and
\begin{equation}
\Psi (z) = \dfrac{\left( \delta(1\,+\,z)^{3(1 + w_{x})} +  3 w_{x}(1+z)^{3 - \delta} \right)}{\delta + 3 w_{x}}.
\label{eq2:2.10}
\end{equation}
 
This model is characterized by the following set of parameters $\Theta=\left\lbrace h,\Omega_b, \Omega_{k},\Omega_{m}, w_{x},\delta \right\rbrace$. 
We~show the best fit values of these parameters in Table \ref{tab:IDE}.
 
\begin{table}[htb]
\centering
\begin{tabular}{lcc}
\hline 
\textit{Parameter} & \textit{CMB+BAO+SNIa} & \textit{CMB+BAO+SNIa+$d_A$+$f_{gass}$+SGL} \\
\hline 
\hline 
h                         & $0.703\pm 0.015$      & $0.7165\pm 0.0092$\\ 
$\Omega_m$     & $0.2963\pm 0.0092$  & $0.2844\pm 0.0049$ \\
$\Omega_k$      & $-0.0048\pm 0.0036$  & $0.0022\pm 0.0029$ \\
$\omega_x$       & $-1.059\pm 0.042$     & $-1.074\pm 0.038$ \\
$\delta$              & $-0.0048\pm 0.0049$ & $-0.0041\pm 0.0036$ \\
$\Omega_b$      & $0.0451\pm 0.0019$  & $0.0428\pm 0.0011$\\
$\chi_{min}^{2}$ & 563.960                      & 771.442 \\
\hline 
\end{tabular} 
\caption{Summary of the best fit values from for IDE model.}
\label{tab:IDE}
\end{table}
 
Is interesting to note that, in both cases, EoS has a phantom behavior at present and the standard model is practically discarded at $1\sigma$ CL. The curvature parameter $\Omega_k$ is positive. We can also notice that the case $\delta=0$ (absence of interaction) is excluded at least to $2\sigma$ CL for the present analysis, where~we can appreciate that for both data sets the transfer of energy is from DM to DE (see Figure \ref{figure:ConDia}).

\subsection{Early Dark Energy model}\label{sec:03.4}
 
In early dark energy (EDE) scenarios, the DE density can be significant at high redshifts. This~may be so if DE fluid tracks the dynamics of the background fluid density \cite{SteinhardtEDE}. Here, we present the EDE model  proposed by \cite{2006JCAP...06..026D}. The FLRW equation for this model is
 
\begin{equation}
 E^{2}(z,\Theta) = \frac{\Omega_{r}(1+z)^{4} +\Omega_{m}(1+z)^{3} + \Omega_{k}(1+z)^{2}}{1 - \Omega_{X}},
 \label{eq2:2.11}
\end{equation}
where $\Omega_{X}$ is given by
 
\begin{eqnarray}
 \Omega_{X} = \frac{\Omega_{X_0} - \Omega_{e}\left[ 1 - (1+z)^{3 w_{0}} \right]  }{\Omega_{X_0} + f(z)} + \Omega_{e} \left[ 1 - (1+z)^{3 w_{0}} \right]\
 \label{eq2:2.12}
 \end{eqnarray}
and
 
\begin{equation}
f(z) = \Omega_{m}(1+z)^{-3w_{0}} + \Omega_{r}(1+z)^{-3w_{0} +1 } + \Omega_{k}(1+z)^{-3w_{0} -1 },
\label{eq2:2.13}
\end{equation}

\noindent such that $\Omega_{X_0} = 1 - \Omega_{m} - \Omega_{k}- \Omega_{r}$ is the current DE density, $\Omega_e$ is the asymptotic early DE density and $w_0$ is the present DE EoE. Here, we have six free parameters 
$\Theta=\left\lbrace h,\Omega_b,\Omega_{k},\Omega_{m}, \Omega_{e},w_{0}\right\rbrace$. The~best fit values of the model parameters are summarized in Table \ref{tab:EDE}.
 
 \begin{table}[htb]
 \centering
 \begin{tabular}{lcc}
 \hline 
  \textit{Parameter} & \textit{CMB+BAO+SNIa} & \textit{CMB+BAO+SNIa+$d_A$+$f_{gass}$+SGL} \\ 
 \hline 
 \hline 
h                         & $0.723\pm 0.019$    &   $0.7154 \pm 0.0099$ \\
$\Omega_m$     & $0.295\pm 0.010$     &  $0.2839 \pm 0.0050 $ \\
$\Omega_k$      & $0.0072\pm 0.0068$ & $ 0.0032 \pm 0.0035$ \\
$\Omega_e$      & $0.043\pm 0.029$     & $0.012  \pm 0.016$  \\ 
$\omega_0$       & $-1.113\pm 0.061$    & $-1.087 \pm 0.040$   \\
$\Omega_b$      & $0.0425\pm 0.0023$ & $0.0429 \pm 0.0012$\\ 
$\chi_{min}^{2}$ &  564.275                    & 771.697  \\
 \hline 
 \end{tabular} 
 \caption{Summary of the best fit values for EDE model.}
 \label{tab:EDE}
 \end{table}
 
\begin{figure*}[htbh]
     \centering
      \includegraphics[width=0.3\textwidth,angle=0]{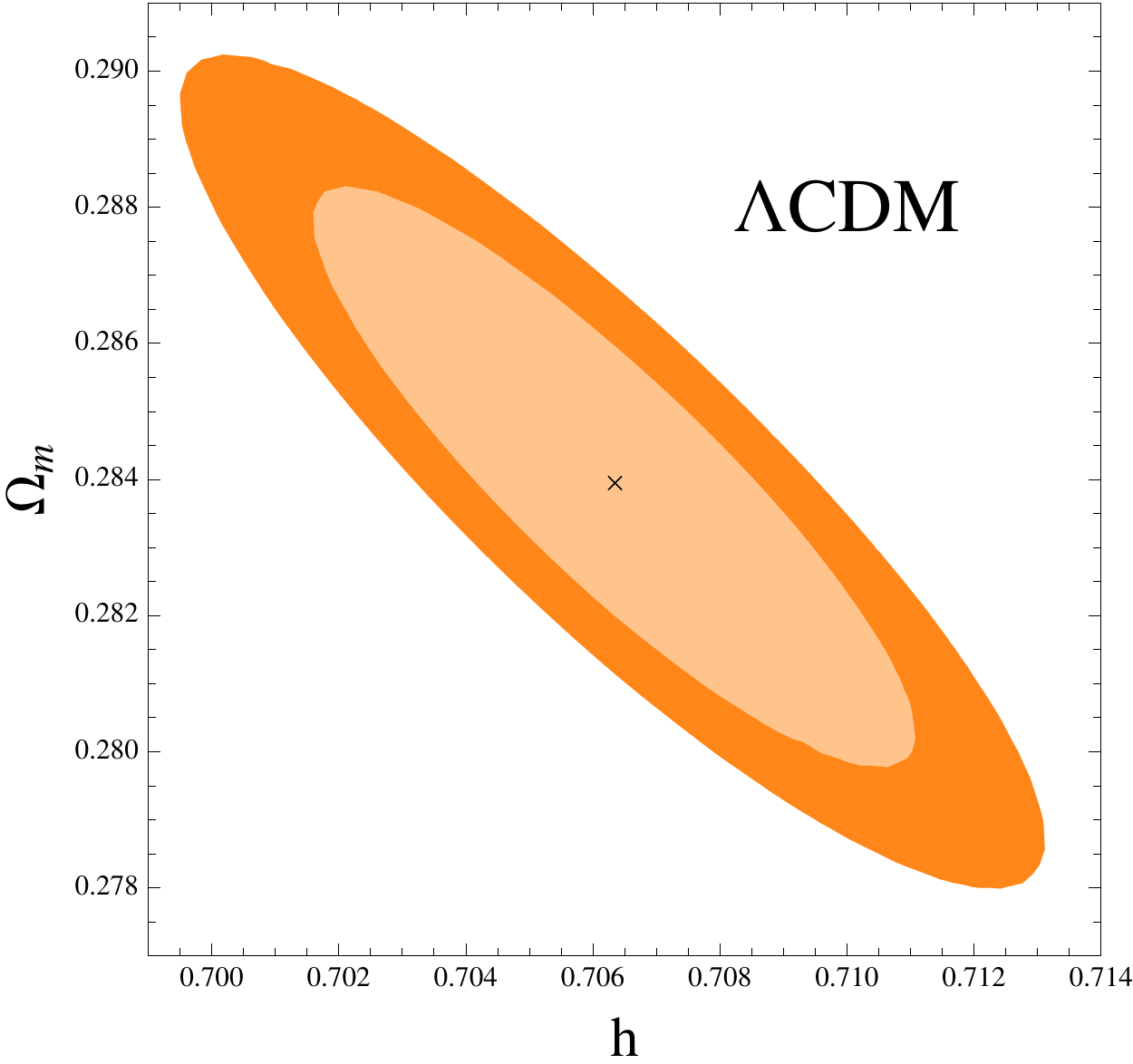}
      \includegraphics[width=0.29\textwidth,angle=0]{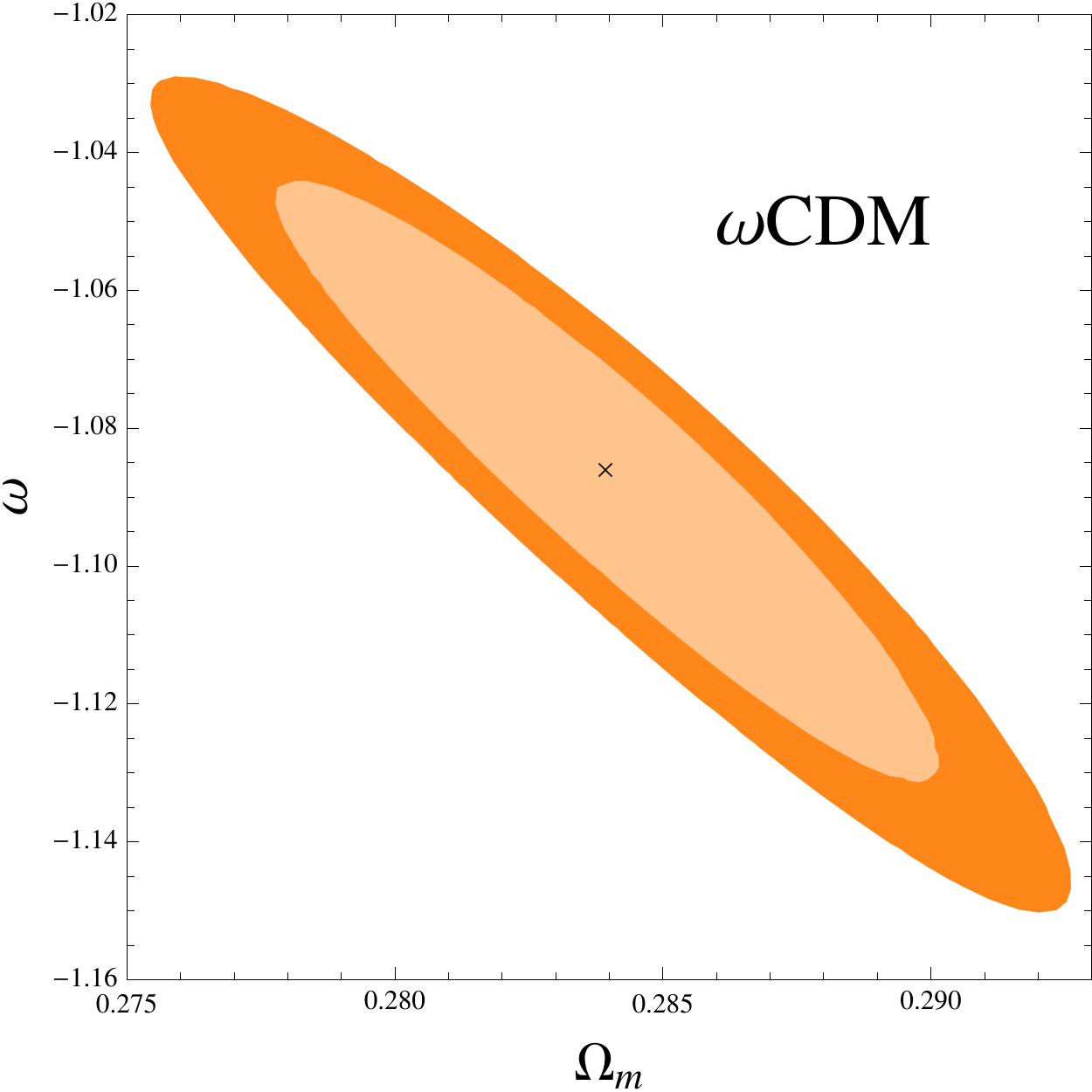}
       \includegraphics[width=0.3\textwidth,angle=0]{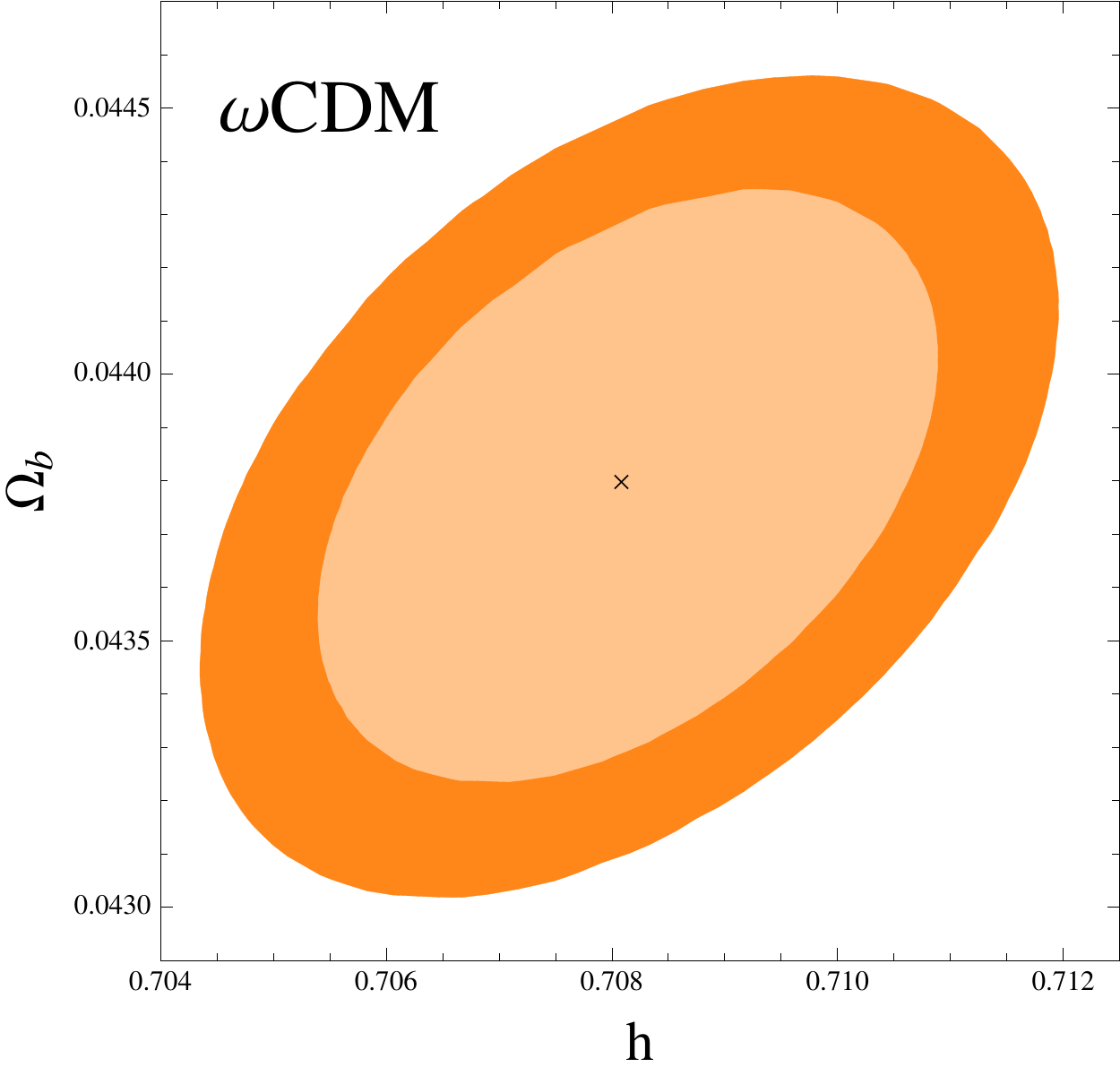}\\
       \includegraphics[width=0.3\textwidth,angle=0]{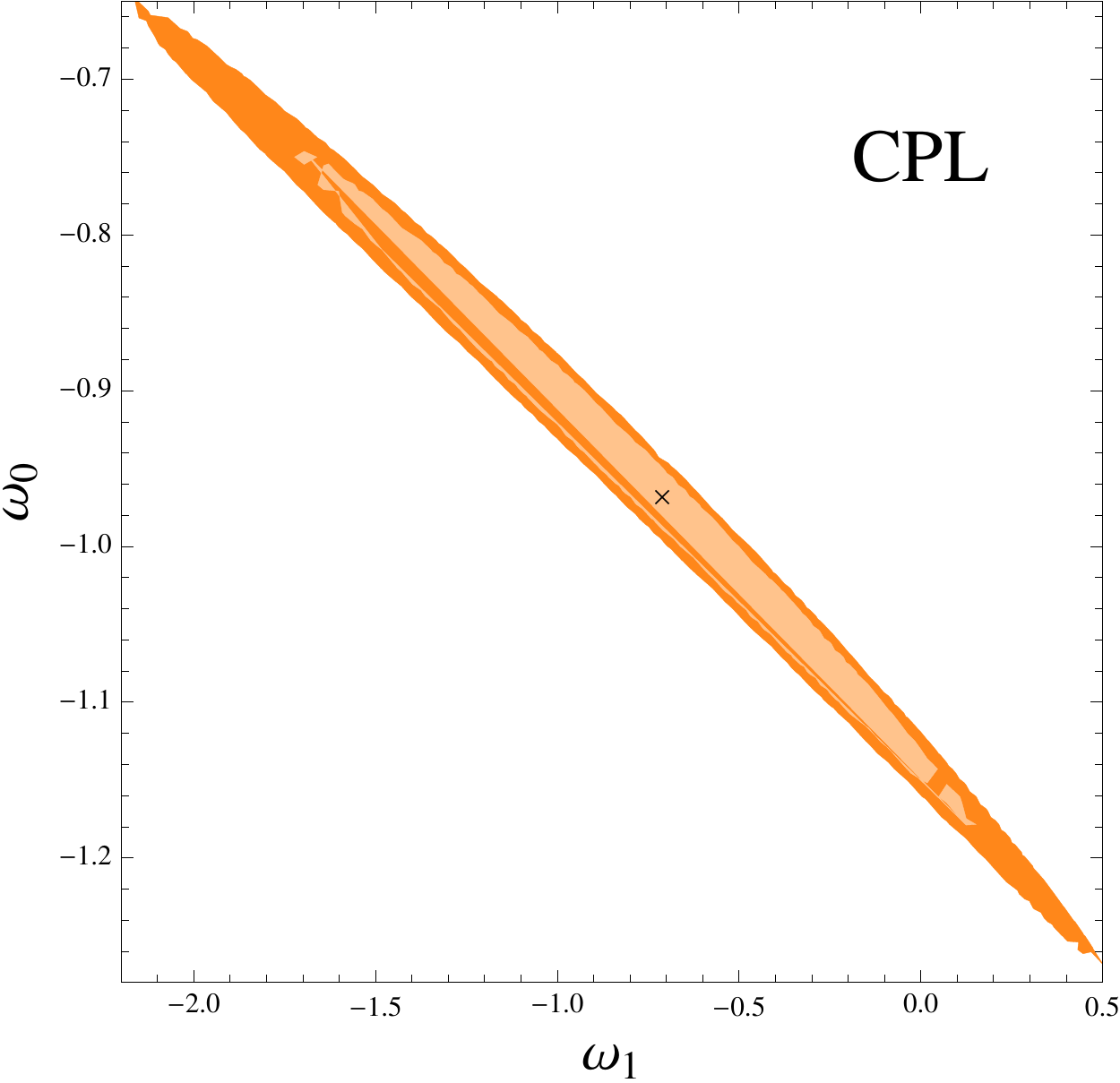}
      \includegraphics[width=0.29\textwidth,angle=0]{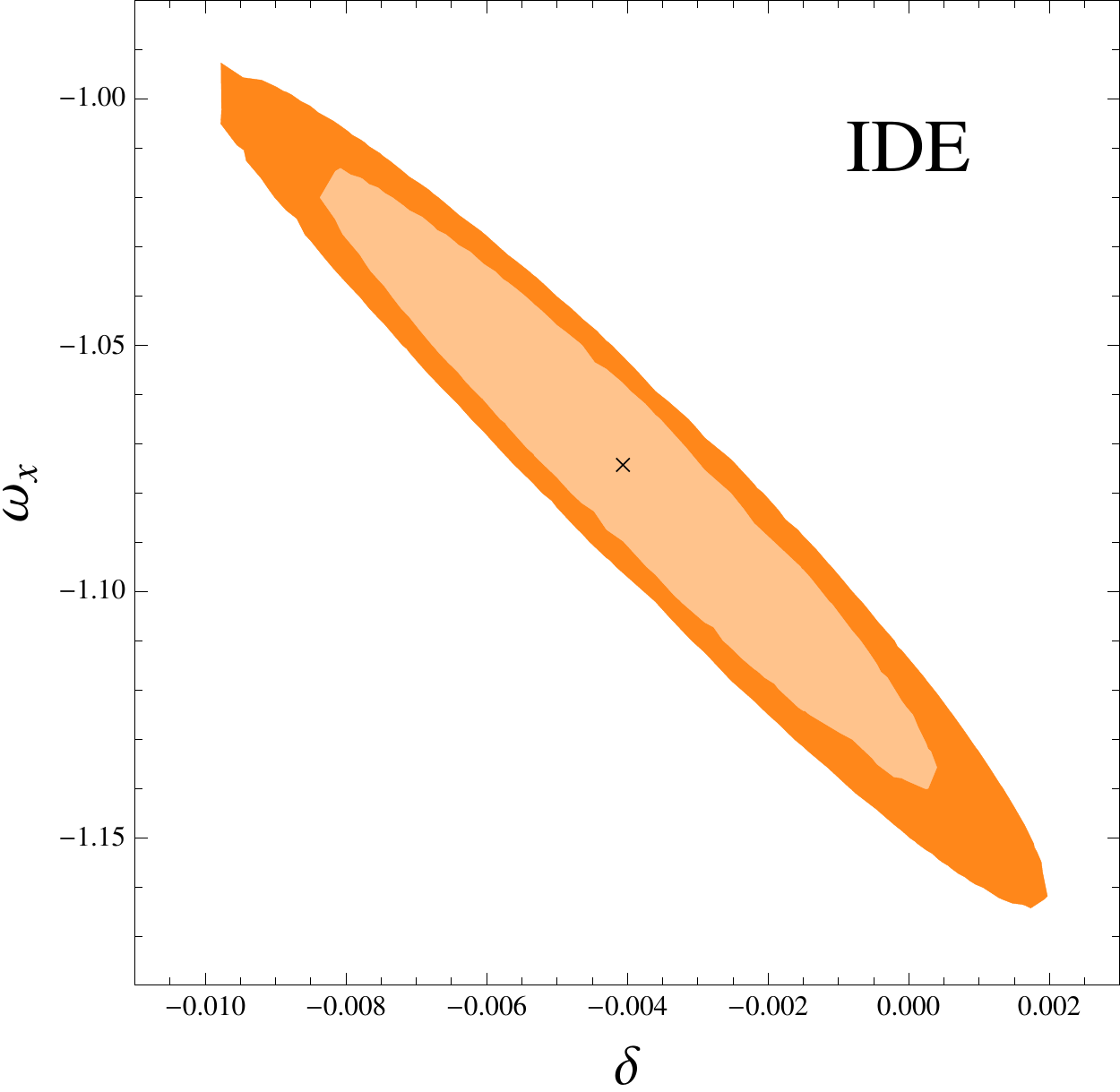}
      \includegraphics[width=0.3\textwidth,angle=0]{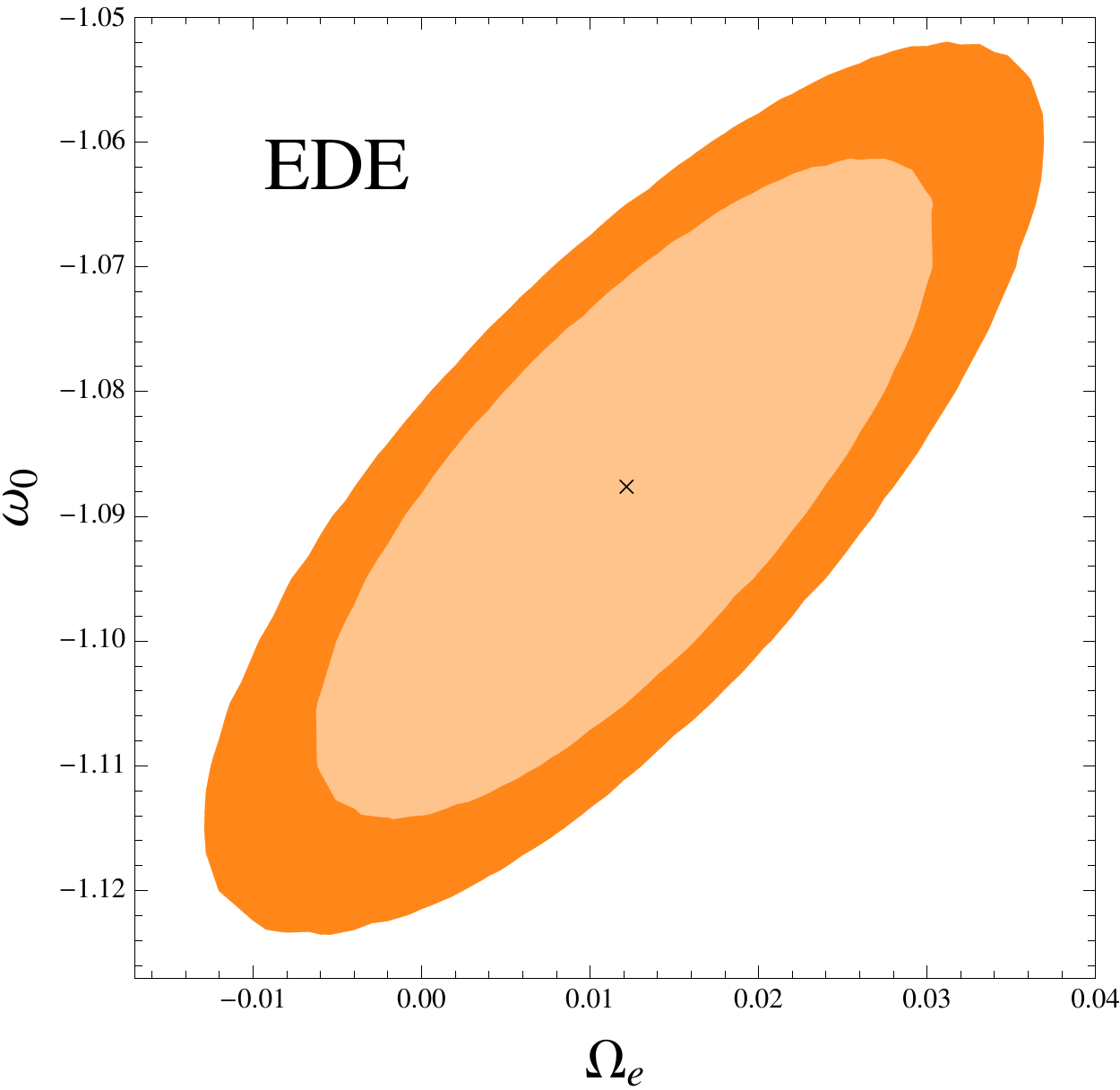}
     \caption{$1\sigma$ and $2\sigma$ two-dimensional CL contours of DE cosmological models discussed, where the main results of the analysis are shown using the combined data sets (CMB+BAO+SNIa+$d_A$+$f_{gass}$+SGL).}
     \label{figure:ConDia}
\end{figure*}

For this model, the EoS  keeps a phantom behavior at the present time and the standard model is discarded at least to $2\sigma$ CL (see Figure \ref{figure:ConDia}). $\Omega_k$ is positive in both cases. 

\subsection{Statistical discrimination models}\label{sec:03.5}

In Table \ref{tab:AIC/BIC}, we present the values for the analysis of the information criterion with respect to the five cosmological models presented above, for used data set, namely CMB + BAO + SNIa + $d_A$ + $f_{gass}$ +~SGL. As we can see, $\Delta AIC$ and $\Delta BIC$ are in favor of  $\omega CDM$ and $\Lambda CDM$, respectively (approximately or less than two), and, hence, these models are in very good agreement with observations, which is also true for CPL, IDE and EDE models only with respect to $\Delta AIC$. For models CPL, IDE and EDE, the~value of $\Delta BIC$ is approximately equal to  seven and therefore, according to this criterion, present less observational support.

\begin{table}
\centering
\begin{tabular}{lcccccc}
\hline 
\hline 
Model                   & $d$ & $\chi^2_{red}$ & \textit{AIC} & $\Delta AIC$ &  \textit{BIC} & $\Delta BIC$ \\ 
\hline 
$\Lambda CDM$  & 4    & 1.060                  & 785.256    & 2.973             & 803.666       & 0.000           \\ 

$\omega CDM$    & 5    & 1.055                  & 782.283    & 0.000             & 805.295       & 1.626            \\ 

\textit{CPL}           &  6   & 1.055                  & 783.481    & 1.198             & 811.096        & 7.430            \\ 

\textit{IDE}            &  6   & 1.055                  & 783.442    & 1.159             & 811.057        & 7.391            \\ 

\textit{EDE}           &  6   & 1.055                  & 783.697    & 1.414             & 811.312        & 7.646            \\ 
\hline 
\end{tabular}
\caption{AIC and BIC analyses for different DE models using the combined analysis data sets (CMB+BAO+SNIa+$d_A$+$f_{gass}$+SGL), where N=737 and $\chi^2_{red} = \chi^2_{min}/\nu$.}
\label{tab:AIC/BIC}
\end{table}

\section{History of the expansion and cosmography}\label{sec:04}

The kinematics of the universe can be described through the Hubble parameter $H(t)$ and its dependence on time, i.e., the deceleration parameter $q(t)$ \cite{Sandage:62} . Following \cite{2012PhyU...55A...2B}, the scale factor $a(t)$ can be expanded in Taylor series around the current time ($t_0$) as:

\begin{equation}
\frac{a(t)}{a(t_0)} = 1 + \frac{H_0}{1!} \left[ t - t_0 \right]  - \frac{q_0}{2!} H^2_0 \left[ t - t_0 \right]^2 + \frac{j_0}{3!} H^3_0 \left[ t - t_0 \right]^3+...,
\label{eq5:5.1}
\end{equation}

\noindent where in general we can have a kinematic description of the cosmic expansion through the set of~parameters:

\begin{equation}
H(t) \equiv \frac{1}{a}\frac{da}{dt};  q(t)\equiv -\frac{1}{a}\frac{d^2 a}{dt^2} H(t)^{-2}; j(t)\equiv\frac{1}{a}\frac{d^3 a}{dt^3} H(t)^{-3},
\label{eq5:5.2}
\end{equation}

\noindent where the last term is know as jerk parameter \emph{j}(\emph{t}). The great advantage of this method is that we can investigate the cosmic acceleration without assuming any modification of gravity theory or DE model, due mainly to its geometric approximation. Although more terms of the series can be analyzed, we are only interested in the first three terms for the present work. The deceleration and jerk parameter are obtained as

\begin{equation}
 q(z) = -1+\frac{(1+z)}{H(z)}\frac{dH(z)}{dz}
\label{eq5:5.3}
\end{equation}
and
\begin{equation}
j(z) = q^2 + \frac{(1+z)^2}{H(z)}\frac{d^2H(z)}{dz^2}.
\label{eq5:5.4}
\end{equation}

The history of expansion is fit through deceleration parameter, which characterize whether the universe is currently accelerated or decelerated 
\begin{equation}
q(z)\equiv - \frac{\ddot{a}(z)}{a(z)H(z)^2}. 
\label{eq5:5.5}
\end{equation}

If $q(z)>0$, $\ddot{a}(z)<0$, then the expansion decelerates as expected due to gravity produced by DM, baryonic matter or radiation. The discovery that the universe today presented an accelerated expansion already has about one decade and a half old \cite{1999ApJ...517..565P,1998AJ....116.1009R}. A simple explanation for this phenomenon is the cosmological constant $\Lambda$, which, however, does not offer a consistent theoretical explanation based on physical foreground. The information about the dynamics of the expansion can be obtained through Equations (\ref{eq5:5.3}) and (\ref{eq5:5.4}), which directly depends on the cosmological model. In~general, if $\Omega_X \neq 0$ is sufficiently large (i.e., $\Omega_X >\Omega_m $), then $q(z)<0$ and $\ddot{a}(z)>0$, which translates into an accelerated expansion as it is shown by the observations. If the accelerated expansion is driven by a new type of fluid, then is important to identify if fluid energy density is constant or dynamic.

\begin{table*}[htbh]
\centering
\begin{tabular}{lcl}
\hline
 Model & $\chi_{red}^{2}$ & Parameters \\
\hline
\hline
$\Lambda$ CDM & 1.11 & $h=0.722 \pm 0.012$, $\Omega_m = 0.2640 \pm 0.0093 $, $\Omega_k =-0.13\pm 0.16$, $\Omega_b=0.0410\pm0.0014$   \\

$\omega$ CDM   & 1.11 & $h = 0.722 \pm 0.012$, $\Omega_m = 0.2685 \pm 0.0093 $, $\Omega_k = -0.14 \pm 0.88 $,  $\omega =-0.99 \pm 0.73$, $\Omega_b=0.0409\pm0.0015$\\

CPL                     & 1.14 & $h = 0.721 \pm 0.011$, $\Omega_m = 0.274 \pm 0.013$, $\Omega_k = -0.5\pm 1.8$,  $\omega_a = -1.5 \pm 2.2$,  $\omega_0 = -0.60 \pm 0.50$, $\Omega_b=0.0411\pm0.0014$ \\ 

IDE                      & 1.14 & $h = 0.721 \pm 0.011$, $\Omega_m =0.274 \pm 0.012$, $\Omega_k = 0.2 \pm 2.5$, $\omega_x = -1.1\pm 3.1$, $\delta = 3.9\pm 13.0$, $\Omega_b=0.0411\pm0.0015$ \\

EDE                     & 1.14 & $h = 0.720 \pm 0.012$, $\Omega_m =0.276 \pm 0.014$, $\Omega_k = 0.3 \pm1.7$, $\omega_0 = -0.8\pm 1.7$, $\Omega_e = -1.1\pm 1.7$, $\Omega_b=0.0412\pm0.0015$ \\
\hline
\end{tabular}
\caption{The best fit values for the free parameters using data from GC ($d_{A}+f_{gas}$).
\label{tab:cls}}
\end{table*}

\begin{figure}[htb]
\centering
\begin{center}
    \includegraphics[height=9cm]{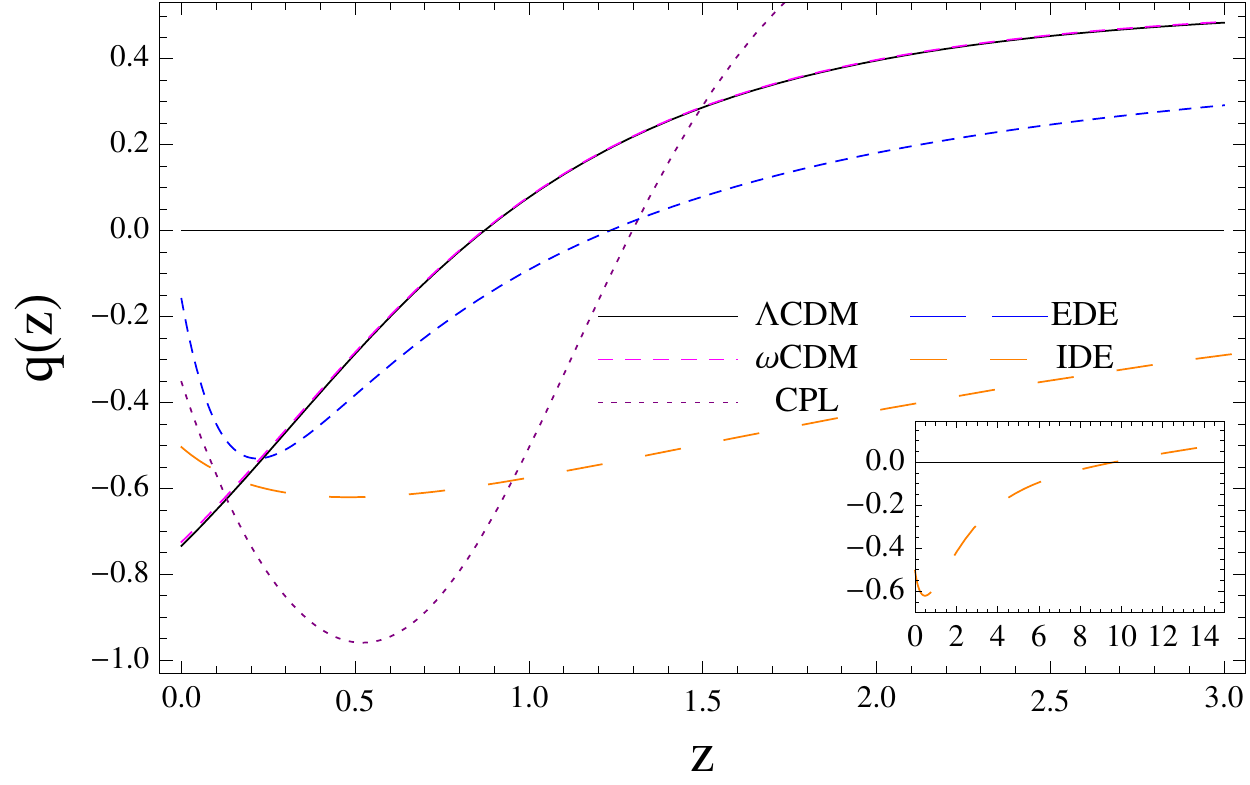}
    \caption{Deceleration parameter vs redshift using only GC data ($d_{A}+f_{gas}$). It is shown the transition decelerated-accelerated ($q(z_t)=0$) and the current value of ($q_0$) ($\Lambda CDM$ ($z_t\sim 0.86$, $q_0=-0.76$), $\omega CDM$ ($z_t\sim 0.86$, $q_0=-0.76$), CPL ($z_t\sim 1.32$, $q_0=-0.35$), IDE ($z_t\sim 9.78$,  $q_0=-0.50$), EDE ($z_t\sim 1.22$, $q_0=-0.17$)).  Notice the strange behavior of the deceleration parameter to later times for models of dynamical DE (CPL, IDE, EDE).}
    \label{fig:qz}
\end{center}
\end{figure}

In the present cosmographic analysis, we use of data from GC ($d_{A}+f_{gas}$), where we can see that these do not provide a tight constraint on curvature and DE parameters, mainly due to the degeneracy presented between these parameters and the large systematic errors of the samples (see~Table \ref{tab:cls}), which can lead to large discrepancies with respect to the standard model. Despite this, we are more interested in the analysis of the behavior of low redshift of each cosmological model with respect to these data sets. Figure \ref{fig:qz} shows the plot of the deceleration parameter $q(z)$ and, as expected, the~models studied give $q(z)<0$ at late times and  $q(z)>0$ at earlier epoch. All cosmological models present a redshift of transition ($z_t$) between the two periods; however, all models of dynamical DE present an interesting behavior of slowing down of acceleration at low redshift (late times), which can be characterized through the change of sign of the parameter $j(z)$ (CPL: $j(z_{low}) \rightarrow 0$, when $z_{low}\sim0.50$; IDE: $j(z_{low})\rightarrow 0$, when $z_{low}\sim0.41$; EDE: $j(z_{low}) \rightarrow 0$, when $z_{low}\sim0.23$). We can interpret $j(z)$ as the slope at each point of $q(z)$, which indicates a change in acceleration. This result is consistent with the one presented by Barrow, Bean and Magueijo \cite{2000MNRAS.316L..41B}, in which arises the possibility of a scenario with accelerated expansion of the universe and that does not imply an eternal accelerated expansion. In~\cite{2012PhyU...55A...2B}, an extensive analysis of this possibility is made (see also\cite{2011CQGra..28l5026G}), which includes a cosmographic analysis like the one presented in the current work. This transient accelerating phase can be also a clear behavior of dynamical DE at low redshift for models with variation of the density of DE over time.

\begin{figure}[htb]
\centering
\begin{center}
    \includegraphics[height=9cm]{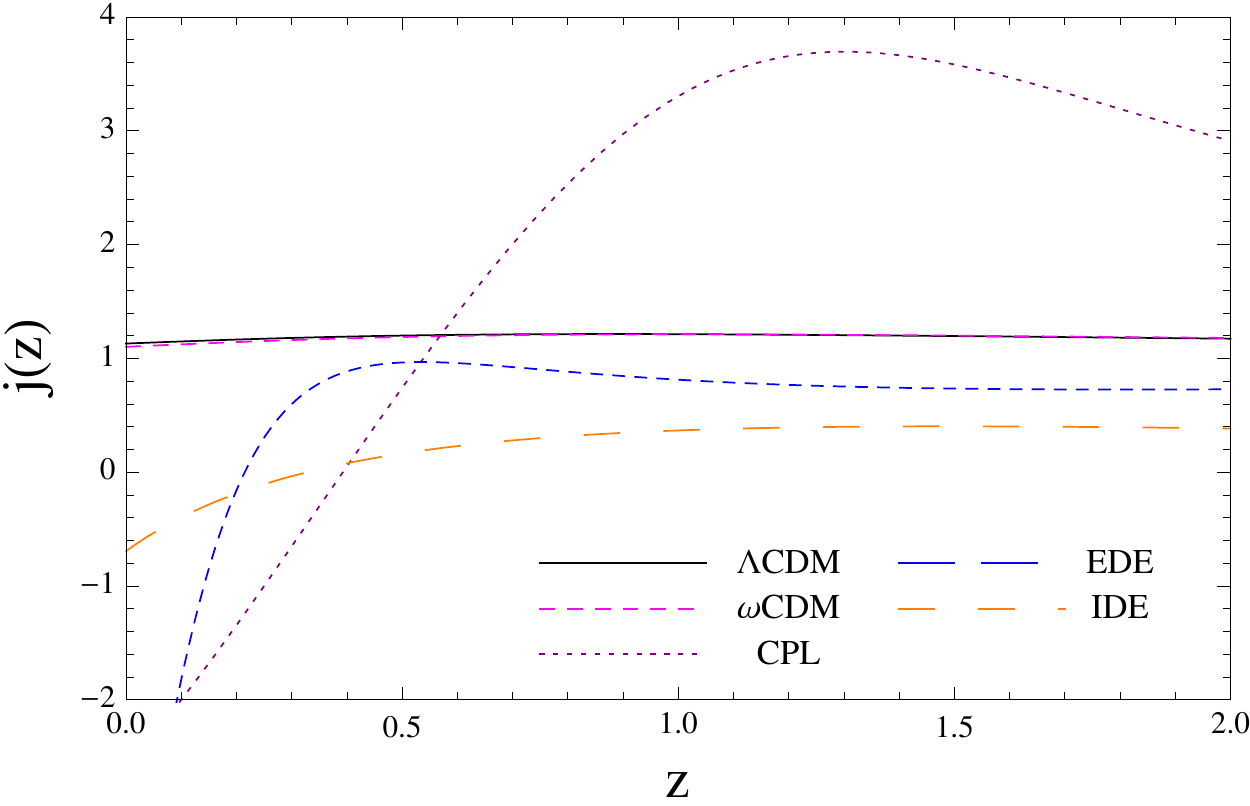}
    \caption{Jerk parameter vs redshift using only GC data ($d_{A}+f_{gas}$). For cosmological models CPL, IDE and EDE, we can observe a strong deviation from $\Lambda CDM$ at present, while for $wCDM$ this does not happen.}
    \label{fig:jz}
\end{center}
\end{figure}

\section{Summary and discussion}
\label{sec:05}

In the present work, we compared alternative cosmological models of DE using data obtained from GC and SGL in addition to more traditional ones, getting the best-fit value of parameters for each one. On the other hand, applying the Akaike and Bayesian information criteria, we determine which of these models is the most favored by current observational data. Our analysis shows that $\omega CDM$ and $\Lambda CDM$ DE models are preferred by $\Delta AIC$ and $\Delta BIC$, respectively. For the first time, we report that the $\omega CDM$ model is favored by observational data at least with $\Delta AIC$; however, the $\Lambda CDM$ model remains the best fit for $\Delta BIC$. In Figure \ref{figure:ConDia}, we can see that $\Lambda CDM$ model is excluded at least $2\sigma$ CL for $\omega CDM$, IDE and EDE models, combining all data sets (see also Tables \ref{tab:WCDM}, \ref{tab:IDE} and \ref{tab:EDE}). Models such as CPL, IDE~and EDE, although they are penalized given their large number of free parameters, have a good fit with the observational data.

On the other hand, we carried out the study of the history of cosmic expansion through the $H(z)$, $q(z)$~and $j(z)$ parameters with data from GC ($d_{A,clusters}+f_{gas}$). We find new evidence showing anomalous behavior of the deceleration parameter q(z) in later times ($z_{low}< 0.5$), suggesting that the expansion of the universe could decelerate in the near future (Figure \ref{fig:qz}), which was pointed out in previous works with SNIa (for CPL \cite{2009PhRvD..80j1301S,2011PhLB..695....1L}), $f_{gas}$ (for CPL and different parameterizations of \textit{w(z)}~\cite{2013MNRAS.433.3534C,2017MNRAS.469...47M}) and BAO (for CPL, IDE and EDE  \cite{2016arXiv160501984B}). Other types of mechanisms were also taken into account to explain this phenomenon (see, for example, \cite{2014arXiv1404.3273P}). This perspective raises the possibility that an accelerated expansion does not imply the eternal expansion, even in the presence of DE \cite{2013IJTP..tmp..453C}. This cosmic slowing down of acceleration only appears in dynamic models of DE (CPL, IDE and EDE), which in principle can be an indication of the need for a scalar field such as quintessence or phantom (see, for example, \cite{2006PhRvL..97h1301C}). Finally, in Figure \ref{fig:jz}, we show the results for jerk parameter $j(z)$ obtained from our kinematic analysis, where we can appreciate a considerable deviation from $\Lambda CDM$ (black curve) in late times ($z < 0.5$) for CPL, IDE and EDE models. A more careful study might give insight into this anomalous behavior, which may also represent a challenge for alternative models to DE including modified gravity models.

As we can see, the fit of observational data acquires slightly larger values of $\chi^2_{min}$ with respect to $\Lambda CDM$, when GC and SGL data are added to the more traditional ones as CMB + BAO + SNIa, which~may be mainly due to their large systematic errors (GC + SGL) (see Tables \ref{tab:LCDM}--\ref{tab:EDE}). However, the~potential of these data sets as cosmological tests is very high, since, for example, the increase in the number of data points and the reduction of systematic errors leads to better constraints in parameters such as DE, which is of fundamental interest for modern cosmology.

\section*{Acknowledgments}
\label{sec:06}

\noindent Alexander Bonilla and Jairo E. Castillo wish to acknowledge to the Universidad Distrital Francisco Jos\'e de Caldas and FIZMAKO group for their academic support. Alexander Bonilla also wishes to thank to the Departamento de F\'isica of the Universidade Federal de Juiz de Fora for his academic support and to Rafael Nunes for constructive and fruitful discussions.

%%%%%%%%%%%%%%%%%%%%%%%%%%%%%%%%%%%%%%%%%%%%%%%%%%%%%%%%%%%%%%%%%%%

\def \aap {A\&A} 
\def \aapr {A\&AR} 
\def \statisci {Statis. Sci.} 
\def \physrep {Phys. Rep.} 
\def \pre {Phys.\ Rev.\ E.} 
\def \sjos {Scand. J. Statis.} 
\def \jrssb {J. Roy. Statist. Soc. B} 
\def \pan {Phys. Atom. Nucl.} 
\def \epja {Eur. Phys. J. A} 
\def \epjc {Eur. Phys. J. C} 
\def \jcap {J. Cosmology Astropart. Phys.} 
\def \ijmpd {Int.\ J.\ Mod.\ Phys.\ D} 
\def \nar {New Astron. Rev.} 

\def \JCAP {JCAP}
\def \araa {ARA\&A}
\def \aj {AJ}
\def \aar {A\&AR}
\def \apj {ApJ}
\def \apjl {ApJL}
\def \apjs {ApJS}
\def \asl {Adv. Sci. Lett.} 
\def \mnras {Mon.\ Non.\ Roy.\ Astron.\ Soc.}
\def \nat {Nat}
\def \pasj {PASJ}
\def \pasp {PASP}
\def \science {Science}

\def \gca {Geochim.\ Cosmochim.\ Acta}
\def \npa {Nucl.\ Phys.\ A}
\def \plb {Phys.\ Lett.\ B}
\def \prc {Phys.\ Rev.\ C}
\def \prd {Phys.\ Rev.\ D.}
\def \prl {Phys.\ Rev.\ Lett.}
\def \rmp {Rev.\ Mod.\ Phys.}

%%%%%%%%%%%%%%%%%%%%%%%%%%%%%%%%%%%%%%%%%%%%%%%%%%%%%%%%%%%%%%%%%%%

\appendix

\section{Appendix A}\label{AppxA}

\subsection{$\beta$-model and triaxial ellipsoids}

\begin{figure*}[htbh]
     \centering
      \includegraphics[width=0.6\textwidth,angle=0]{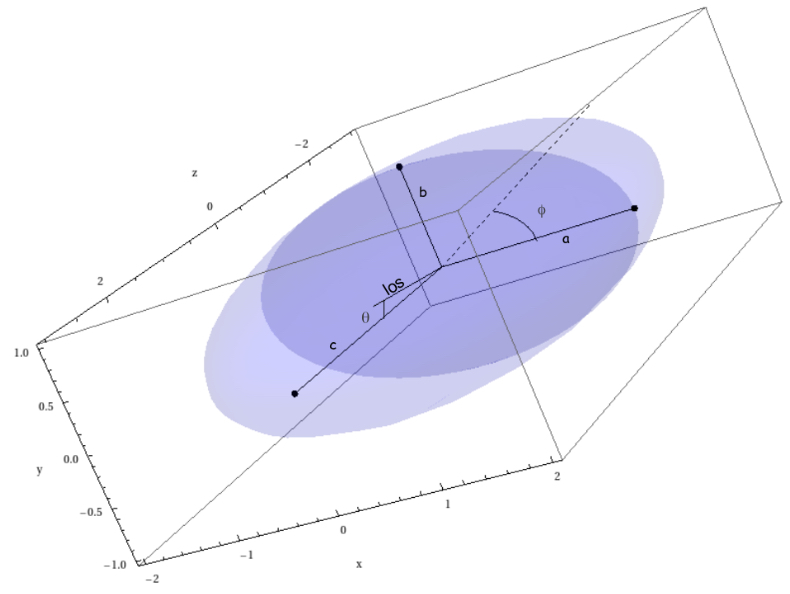}
      \includegraphics[width=0.6\textwidth,angle=0]{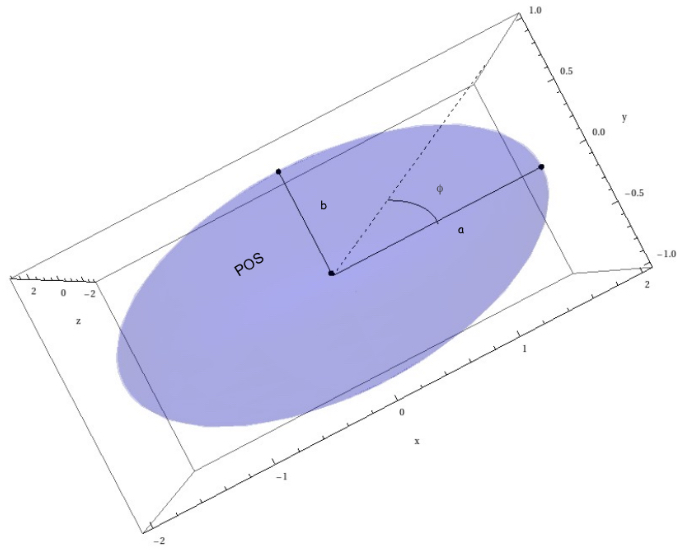}
     \caption{Ellipsoid coefficients a, b and c, with the \textit{los} making an angle $\theta$ with the z-axis (up). View of the \textit{pos} with the \textit{los} oriented along the z-axis (down).}
     \label{figure:Ellp}
\end{figure*}

In the distribution described by an ellipsoidal triaxial $\beta$-model, the electron density of the intracluster gas is assumed to be constant on a family of similar, concentric, coaxial ellipsoids. In a coordinate system relative to GC, the electron density distribution is

\begin{equation}
n_e = n_{e0} \left( 1 + \frac{\sum_{i=1}^3v_i^2x_{i,int}^2}{r_c^2}\right)^{-3\beta/2}
\end{equation}

\noindent where $x_{i,int}$ is the intrinsic orthogonal coordinate system centred on GC's barycenter, $r_c$ is characteristic length scale distribution at core radius, $v_i$ is the inverse of the corresponding core core radius, $n_{e0}$ is the central electron density. if we take the axial ratios $e_1\equiv v_1/v_2$, $e_2\equiv v_2/v_3$, $r_{c3}=r_c/v_3$ and and taking into account that

\begin{equation}
\frac{x}{a} + \frac{y}{b} + \frac{z}{c} = r_{ellp}
\end{equation}

\noindent such that $x_1=x$, $x_2=y$, $x_3=z$ and $v_1=a^{-1}$, $v_2=b^{-1}$, $v_3=c^{-1}$  (see Fig. \ref{figure:Ellp}), we can obtain

 \begin{equation}
n_e = n_{e0} \left( 1 + \frac{e_1^2x_{1,int}^2 + e_2^2x_{2,int}^2 + x_{3,int}^2}{r_c^2}\right)^{-3\beta/2}
\end{equation} 

\noindent with 

\begin{equation}
\beta = \frac{\nu m_p \sigma_v^2}{k_BT_e}.
\end{equation}

Then the electron density distribution is described by five parameters in a ellipsoidal triaxial $\beta$-model: $n_{e0}$, $\beta$, $e_1$, $e_2$ and $r_{c3}$. 

The projection along the \textit{los} of the electron density distribution, to a generic power \textit{m} in the observer coordinate system is given by

\begin{equation}
\int_{los} n_e^m (l) dl = n_{e0}^m \sqrt{\pi}  \frac{\Gamma(3m\beta-1/2)}{\Gamma(3m\beta2)} \frac{d_A\theta_3}{\sqrt{h}} \left( 1 + \frac{\theta_1^2 + e_{proj}^2 \theta_2^2}{\theta_{c,proj}^2} \right)^{1/2-3\beta/2}
\end{equation}

\noindent where $d_A$ is the angular diameter distance in a \textit{FRW} universe, $\theta_i\equiv x_{i,obs}/d_A$ $e_{proj}$ is the projected angular position on the plane of the sky (\textit{pos}) of the intrinsic orthogonal coordinate system $x_{i,obs}$ and \textit{h} is a function of the GC shape and orientation:

\begin{equation}
h = e_1^2 sin^2 \theta_{Eu} sin^2 \varphi_{Eu} + e_2^2 sin^2 \theta_{Eu} cos^2 \varphi_{Eu} + cos^2 \theta_{Eu}, 
\end{equation}

\noindent such that $\theta_{Eu}$ and  $\varphi_{Eu}$ are the Euler angles in the GC coordinate system (see Fig. \ref{figure:Ellp}) and 

\begin{equation}
\theta_{c,proj} \equiv \theta_{c3} \left( \frac{e_{proj}}{e_1e_2}\right) ^{1/2} h^{1/4}.
\end{equation}

If we assume that the intracluster medium is described by an isothermal triaxial $\beta$-model distribution with m=1 we obtain

\begin{equation}
\Delta T_{sz} = \Delta T_{sz0} \left( 1 + \frac{\theta_1^2 + e_{proj}^2 \theta_2^2}{\theta_{c,proj}^2} \right)^{1/2-3\beta/2} 
\end{equation}

\noindent where $\Delta T_{sz0}$ is the central temperature decrement of SZ effect, which is given by

\begin{equation}
\Delta T_{sz0} \equiv T_{cmb} f(\nu,T_e) \frac{k_BT_e}{m_ec^2} n_{e0} \sqrt{\pi} \frac{d_A \theta_{c,proj}}{h^{4/3}}\sqrt{\frac{e_1e_2}{e_{proj}}} g\left( \frac{\beta}{2}\right) 
\end{equation}

\noindent and

\begin{equation}
g\left( \alpha \right) \equiv \frac{\Gamma(3\alpha-1/2)}{\Gamma(3\alpha)}.  
\end{equation}

$e_{proj}$ is the axial ratio of the major to minor axes of the observed projected isophotes and $\theta_{c,proj}$ is the projection on the (\textit{pos}).

On the other hand, the X-Ray surface brightness for intracluster medium with m=2, is given by

\begin{equation}
S_x = S_{x0} \left( 1 + \frac{\theta_1^2 + e_{proj}^2 \theta_2^2}{\theta_{c,proj}^2} \right)^{1/2-3\beta/2} 
\end{equation}

\noindent where the central surface brightness $S_{x0}$ is

\begin{equation}
S_{x0} \equiv  \frac{\Lambda_{eH}(\mu_e/\mu_H)}{4\sqrt{\pi}(1+z)^4} n_{e0}  \frac{d_A \theta_{c,proj}}{h^{4/3}}\sqrt{\frac{e_1e_2}{e_{proj}}} g\left( \beta\right), 
\end{equation}

\noindent with $\mu_i \equiv \rho/(n_im_p)$ the molecular weight.

\subsection{Galaxy clusters data}

\begin{figure*}[htbh]
     \centering
      \includegraphics[width=0.9\textwidth,angle=0]{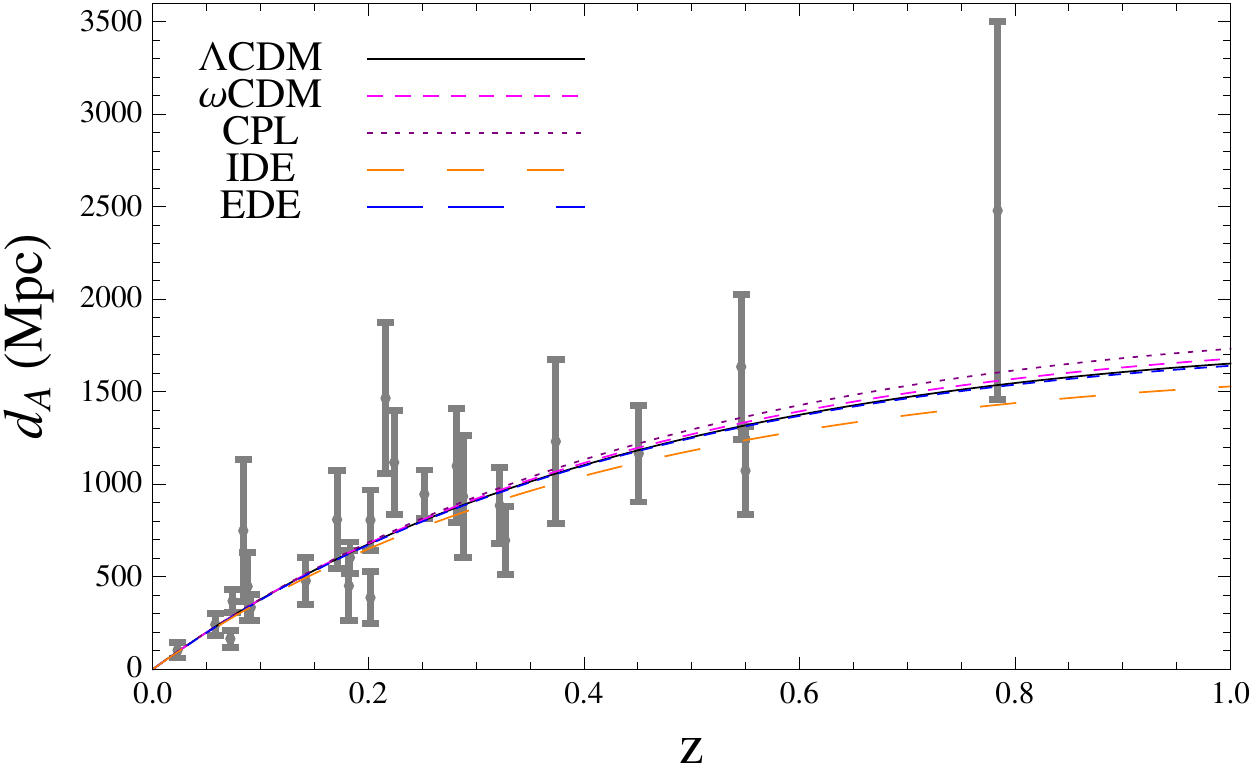}
     \caption{Angular diameter distance vs redshift for different models with the best fit values from joint analysis (CMB+BAO+SNIa+$d_A$+ $f_{gass}$+SGL) and 25 data set from De Filippis et al. (Gray) \cite{DeF05}.}
     \label{figure:dA}
\end{figure*}

Table \ref{tab:Dc} shows us the experimental cosmological distance with triaxial symmetry from De Filippis et al. obtained by the method S-Z/X-Ray \cite{DeF05}. Column 1 shows the cluster identification name, column 2 give the correspond redshift, column 3 is gas temperature,  column 4 is central temperature decrement, column 5 is the term of dependence with frequency with relativistic corrections and column 6 show us the experimental cosmological distance. Fig \ref{figure:dA}. show us the angular diameter distance vs reshift and the data sample from De Filippis et al.

\begin{table*}
\centering
\begin{tabular}{lccccc}
\hline 
\hline 
Cluster                    & $z_i$      & $k_BT_e$(keV)       & $\Delta T_{sz0}$($\mu K$) & $f(\nu,T_e)$ &  $D_{c}|^{ell}_{exp}$ (Mpc)  \\ 
\hline
 MS 1137.5+6625   & 0.784      & $5.7^{+1.3}_{-0.7}$       & $-818^{+98}_{-113}$    & 2.00 &  $2479\pm1023$  \\ 
MS 0451.6-0305     & 0.550      & $10.4^{+1.0}_{-0.8}$     & $-1431^{+98}_{-105}$  & 1.87 &  $1073\pm238$    \\ 
Cl 0016+1609         & 0.546      & $7.55^{+0.72}_{-0.58}$ & $-1242\pm105$            & 1.89 &  $1635\pm391$     \\ 
RXJ1347.5-1145	& 0.451       & $9.3^{+0.7}_{-0.6}$       & $-3950\pm350$            & 1.91 &  $1166\pm262$     \\ 
A 370 		             & 0.374       & $6.6^{+0.7}_{-0.5}$       & $-785\pm118$               & 1.96 &  $1231\pm441$     \\ 
MS 1358.4+6245	& 0.327       & $7.48^{+0.50}_{-0.42}$ & $-784\pm90$                & 1.88 &  $ 697\pm183$      \\ 
A 1995		             & 0.322       & $8.59^{+0.86}_{-0.67}$ & $-1023^{+83}_{-77}$     & 1.91 &  $885\pm207$        \\ 
A 611		             & 0.288       & $6.6\pm0.6$                  & $-853^{+120}_{-140}$   & 1.76 &  $934\pm331$       \\ 
A 697 		             & 0.282       & $9.8\pm0.7$                  & $-1410^{+160}_{-180}$ & 1.89 &  $1099\pm308$      \\ 
A 1835 		             & 0.252       & $8.21^{+0.19}_{-0.17}$ & $-2502^{+150}_{-175}$ & 1.93 &  $946\pm131$        \\
A 2261 		             & 0.224       & $8.82^{+0.37}_{-0.32}$ & $-1697\pm200$             & 1.87 &  $1118\pm283$       \\ 
A 773 		             & 0.216       & $9.29^{+0.41}_{-0.36}$ & $-1260\pm160$             & 1.76 &  $1465\pm407$       \\ 
A 2163 		             & 0.202       & $12.2^{+1.1}_{-0.7}$     & $-1900\pm140$             & 1.90 &  $806\pm163$         \\ 
A 520 		             & 0.202       & $8.33^{+0.46}_{-0.40}$ & $-662\pm95$                 & 1.93 &  $387\pm141$         \\ 
A 1689 		             & 0.183       & $9,66^{+0.22}_{-0.20}$ & $-1729^{+105}_{-120}$ & 1.86 &  $604\pm84$           \\ 
A 665 		             & 0.182       & $9.03^{+0.35}_{-0.31}$ & $-728\pm150$               & 1.87 &  $451\pm189$          \\ 
A 2218 		             & 0.171       & $7.05^{+0.22}_{-0.21}$ & $-731^{+125}_{-100}$   & 1.95 &  $809\pm263$          \\ 
A 1413 		             & 0.142       & $7.54^{+0.17}_{-0.16}$ & $-856\pm110$               & 1.88 &  $478\pm126$          \\ 
A 2142 		             & 0.091       & $7.0\pm0.2$                  & $-437\pm25$                 & 1.87 &  $335\pm70$           \\ 
A 478 		             & 0.088       & $8.0\pm0.2$                  & $-375\pm28$                 & 1.91 &  $448\pm185$          \\ 
A 1651 		             & 0.084       & $8.4\pm0.7$                  & $-247\pm30$                 & 1.75 &  $749\pm385$          \\ 
A 401 		             & 0.074       & $6.4\pm0.2$                  & $-338\pm20$                 & 1.78 &  $369\pm62$           \\ 
A 399 		             & 0.072       & $9.1\pm0.4$                  & $-164\pm21$                 & 1.81 &  $165\pm45$           \\ 
A 2256 		             & 0.058       & $9.7\pm0.8$                  & $-243\pm29$                 & 1.96 &  $242\pm61$            \\ 
A 1656 		             & 0.023       & $6.6\pm0.2$                  & $-302\pm48$                 & 1.96 &  $103\pm42$           \\

\hline 
\end{tabular}
\caption{Galaxy  Cluste data set from De Filippis et al. for 25 data point (S-Z/X-Ray) \cite{DeF05}.}
\label{tab:Dc}
\end{table*}

\section{Appendix B}\label{AppxB}

%\appendix

\subsection{SNIa}

We use the Union $2.1$ compilation, which contains a sample of 580 data points. We can get the luminosity distance through the relation $d_L(z)=(1+z)^2d_A(z)$, then to fit cosmological model by minimizing the $\chi^2$ value defined by

\begin{equation}
\chi_{SNIa}^2 = \textsf{A} - \frac{\textsf{B}^2}{\textsf{C}}
\end{equation}

\noindent where

\begin{eqnarray*}
\textsf{A} = \sum_{i=1}^{580}\frac{
[\mu_{th}(z_{i},p_i)-\mu_{obs}(z_{i})]^2 }{\sigma_{\mu_i}^2},\\
\end{eqnarray*}

\begin{equation}
\textsf{B} = \sum_{i=1}^{580}\frac{
\mu_{th}(z_{i},p_i)-\mu_{obs}(z_{i}) }{\sigma_{\mu_i}^2},
\end{equation}

\begin{eqnarray*}
\textsf{C} = \sum_{i=1}^{580}\frac{1}{\sigma_{\mu_i}^2},
\end{eqnarray*}

\noindent where  $\mu(z)\equiv 5\log_{10}[d_L(z)/\texttt{Mpc}]+25$ is the
theoretical value of the distance modulus, and we have marginalized over the nuisance
parameter $\mu_0$ and $\mu_{obs}$.

\subsection{CMB}

A standar observational test is the angular scale of sound horizon ($r_s$) at time of decoupling ($z_{cmb}\sim 1090$), which is encoded in the location  of the first peak of the CMB power spectrum $l^{TT}_{1}$. We include CMB information of Planck 13 data \cite{planckXVI}, whose minimization is given by

\begin{equation}
 \chi^2_{CMB} = X_{Planck13}^TC_{cmb}^{-1}X_{Planck13},
 \label{eq3:3.1}
\end{equation}

\noindent such that

\begin{equation}
 X _{Planck13}=\left(
 \begin{array}{c}
 l_A - 301.57 \\
 R - 1.7407 \\
\omega_b  - 0.02228
\end{array}\right),
 \label{eq3:3.2}
\end{equation}

\noindent where $\omega_b = \Omega_b h^2$. Here $l_A$ is the "acoustic scale" defined as

\begin{equation}
l_A = \frac{\pi d_A(z_{cmb})(1+z_{cmb})}{r_s(z_{cmb})},
 \label{eq3:3.3}
\end{equation}

\noindent where $d_A(z_{cmb})$ is the angular diameter distance and $z_{cmb}$ is the redshift of decoupling given by \cite{husugi},

\begin{equation}
z_{cmb} = 1048[1+0.00124(\Omega_b h^2)^{-0.738}]
[1+g_1(\Omega_{m}h^2)^{g_2}],
 \label{eq3:3.4}
\end{equation}

\begin{equation}
g_1 = \frac{0.0783(\Omega_b h^2)^{-0.238}}{1+39.5(\Omega_b
h^2)^{0.763}},
 g_2 = \frac{0.560}{1+21.1(\Omega_b h^2)^{1.81}},
  \label{eq3:3.5}
\end{equation}

The ``shift parameter'' $R$ is defined as \cite{BET97}

\begin{equation}
R = \frac{\sqrt{\Omega_{m}}}{c} d_A (z_{cmb}) (1+z_{cmb}).
 \label{eq3:3.6}
\end{equation}

$C_{cmb}^{-1}$ in Eq. (\ref{eq3:3.1}) is inverse covariance matrix for  ($R, l_A, \omega_b$), which to Planck 13 data is:

\begin{equation}
C_{cmb^{Planck13}}^{-1} = \sigma_{i} \sigma_{j} C_{NorCov_{i,j}}, 
 \label{eq3:3.9}
\end{equation}

where $\sigma_{i} = \left( 0.18, 0.0094, 0.00030  \right)$ and normalized covariance matrix is:

\begin{equation}
C_{NorCov_{i,j}} = \left(
\begin{array}{ccc}
1.0000 & 0.5250 & -0.4235\\
0.5250 &  1.0000 & -0.6925\\
-0.4235 & -0.6925 & 1.0000
\end{array}\right).
 \label{eq3:3.10}
\end{equation}

This test contributes with 3 data points to the statistical analysis.

\subsection{BAO}

The large scale correlation function measured from SDSS, includes a peak which was identified with the expanding spherical wave of baryonic perturbations from acoustic oscillations at recombination, whose current comoving scale corresponds to $150Mpc$. The expected BAO scale depends on the scale of the sound horizon at recombination and on transverse and radial scales at the mean redshift of galaxies in the survey. To obtain constraints on cosmological model we begin with $\chi^2$ for WiggleZ BAO data
\cite{Blake et al.(2011)}, which is given by

\begin{equation}
\chi^2_{\scriptscriptstyle WiggleZ} =
(\bar{A}_{obs}-\bar{A}_{th})C_{\scriptscriptstyle
WiggleZ}^{-1}(\bar{A}_{obs}-\bar{A}_{th})^T,
\end{equation}
\noindent where $\bar{A}_{obs} = (0.447, 0.442, 0.424)$ is
data vector at $z=(0.44,0.60,0.73)$ and $\bar{A}_{th}(z,p_i)$ is
given by \cite{2005ApJ...633..560E}

\begin{equation}
\bar{A}_{th}=D_V(z) \frac{\sqrt{\Omega_m H_0^2}}{cz},
\end{equation}

\noindent in which $D_V(z)$ is the distance scale defined as

\begin{equation}
D_V(z) = \frac{1}{H_0}\left[ (1+z)^2 d_A (z)^2
\frac{cz}{E(z)}\right]^{1/3}.
 \end{equation}

\noindent Here, $d_A(z)$ is the angular
diameter distance. Additionally, $C_{\scriptscriptstyle WiggleZ}^{-1}$ is the inverse
covariance matrix for the WiggleZ data set given by

\begin{equation}
C_{\scriptscriptstyle WiggleZ}^{-1} = \left(
\begin{array}{ccc}
1040.3 & -807.5   & 336.8    \\
-807.5  & 3720.3  & -1551.9 \\
336.8   & -1551.9 & 2914.9
\end{array}\right).
\end{equation}

Similarly, for the SDSS DR7 BAO distance measurements,
$\chi^2$ can be expressed as \cite{2010MNRAS.401.2148P}

\begin{equation}
\chi^2_{\scriptscriptstyle SDSS} =
(\bar{d}_{obs}-\bar{d}_{th})C_{\scriptscriptstyle
SDSS}^{-1}(\bar{d}_{obs}-\bar{d}_{th})^T,
 \label{eq3:3.19}
\end{equation}

\noindent where $\bar{d}_{obs} = (0.1905,0.1097)$ is the
data points at $z=0.2$ and $z=0.35$. Here, $\bar{d}_{th}(z_d,p_i)$
denotes the distance ratio

\begin{equation}
\bar{d}_{th} = \frac{r_s(z_d)}{D_V(z)},
 \label{eq3:3.20}
\end{equation}

\noindent in which $r_s(z)$ is the comoving sound horizon
given by

\begin{equation}
 r_s(z) = c \int_z^\infty \frac{c_s(z')}{H(z')}dz',
  \label{eq3:3.13}
 \end{equation}

\noindent and $c_s(z)$ is the sound speed

\begin{equation}
c_s(z) = \frac{1}{\sqrt{3(1+\bar{R_b}/(1+z)}},
 \label{eq3:3.14}
\end{equation}

\noindent with $\bar{R_b} = 31500
\Omega_{b}h^2(T_{CMB}/2.7\rm{K})^{-4}$ and $T_{CMB} = 2.726K$. The
redshift $z_{drag}$ at the baryon drag epoch is fitted with the
formula proposed in \cite{1998ApJ...496..605E},

\begin{equation}
z_{drag} =
\frac{1291(\Omega_{m}h^2)^{0.251}}{1+0.659(\Omega_{m}h^2)^{0.828}}[1+b_1(\Omega_b
h^2)^{b_2}],
 \label{eq3:3.15}
\end{equation}

\noindent where

\begin{equation}
 b_1 = 0.313(\Omega_{m}h^2)^{-0.419}[1+0.607(\Omega_{m}h^2)^{0.674}]
 \end{equation}

\noindent and

\begin{equation}
 b_2 = 0.238(\Omega_{m}h^2)^{0.223}.
\end{equation}

Here $C_{\scriptscriptstyle SDSS}^{-1}$  is the
inverse covariance matrix for the SDSS data set given by

\begin{equation}
C_{\scriptscriptstyle SDSS}^{-1} = \left(
\begin{array}{cc}
30124 & -17227\\
-17227 & 86977
\end{array}\right).
 \label{eq3:3.21}
\end{equation}

For the 6dFGS BAO data \cite{2011MNRAS.416.3017B}, there is only
one data point at $z=0.106$, the $\chi^2$ is easy to compute
\begin{equation}
\chi^2_{\scriptscriptstyle 6dFGS} =
\left(\frac{d_z-0.336}{0.015}\right)^2.
 \label{eq3:3.22}
\end{equation}

Additionally, we include measures from the Main Galaxy
Sample of Data Release 7 of Sloan Digital Sky Survey (SDSS-MGS)
\cite{bao3} $r_s/D_V(0.57)=0.0732 \pm 0.0012$. Then, the total
$\chi^2_{BAO}$ is given by

\begin{equation}
\chi_{BAO}^{2} = \chi^{2}_{WiggleZ} + \chi^{2}_{SDSS} + \chi^{2}_{6dFGS} + \chi^{2}_{SDSS-MGS}
\end{equation}

\end{document}